\documentclass[10pt]{article}

\usepackage[T1]{fontenc}
\usepackage[utf8]{inputenc}
\usepackage[margin=1in]{geometry}
\usepackage{microtype}

\usepackage{amsmath,amssymb,amsfonts,amsthm,mathtools}

\usepackage{graphicx}
\graphicspath{{images/}}
\usepackage{booktabs}
\usepackage{multirow}
\usepackage{array}
\usepackage{tabularx}
\usepackage{longtable}
\usepackage{rotating}
\usepackage{float}

\usepackage{xcolor}
\usepackage{enumitem}

\usepackage[numbers,sort&compress]{natbib}
\usepackage{url}
\usepackage[hidelinks]{hyperref}

\theoremstyle{definition}

\theoremstyle{remark}

\usepackage{algorithm}
\usepackage{algpseudocode}

\usepackage{mathrsfs}      
\usepackage{bbm}           
\usepackage{dsfont}        
\usepackage{subcaption}
\usepackage{color,soul}    
\usepackage{placeins}      
\usepackage{authblk}

\title{Reconstruction of Enhanced Causal Omnidirectional Network (RECON)}

\title{Reconstruction of Enhanced Causal Omnidirectional Network (RECON)}
\author[1]{Praveen Niranda}
\author[2]{Peter T. McKenney}
\author[1]{Guifang Fu\thanks{Corresponding author. Email: gfu@binghamton.edu}}
\affil[1]{Department of Mathematics and Statistics, Binghamton University, Binghamton, NY 13902}
\affil[2]{Department of Immunology and Microbial Disease, Albany Medical College, Albany, NY 12208}
\date{}
\date{}

\begin{document}



\maketitle

\begin{abstract}
Learning a dynamical system and reconstructing the underlying regulatory network from $p$ discretely observed state trajectories remain challenging problems. Existing approaches often produced a large number of spurious edges and suffered from several methodological limitations. We propose a new approach, Reconstruction of Enhanced Causal Omnidirectional Network (RECON), that leverages an integral-based additive nonparametric ODE model to reconstruct regulatory networks from $p$ time-course data. RECON incorporates five methodological advances. First, it incorporates a new data-driven edge selection procedure that substantially reduces spurious edges while preserving true regulatory edges. Second, it reconstructs an omnidirectional network that captures causal regulatory relationships rather than merely statistical associations or noise artifacts. Third, it substantially broadens the applicability of standard ODE-based approaches by accommodating both dense regular and sparse irregular longitudinal sampling scenarios. Fourth, it models both node trajectories and edge regulatory effects as time-varying functions, emphasizing a dynamic regulatory network. Fifth, it reconstructs a signed and weighted regulatory network and provides comprehensive network interpretation through two-way direction, activatory/inhibitory indicator, and strength, together with keystone node identification and topological structure. Across five simulation studies, RECON consistently outperforms GRADE by removing nearly all spurious edges while retaining nearly all true regulatory edges, resulting in highly accurate network reconstruction. In the most challenging scenario, the number of spurious edges is reduced from 239 to 0. The application to a longitudinal gut microbiota dataset from allogeneic hematopoietic cell transplantation patients reveals distinct regulatory dynamics before and after transplantation, providing new biological insights that cannot be inferred from abundance measurements alone. To the best of our knowledge, this is the first study to reconstruct a causal omnidirectional network from this dataset.
\end{abstract}


\section{Introduction}\label{sec:intro}

The human microbiota consists of diverse taxa that form complex ecological networks~\citep{barabasi_network_2004}. These microbial communities, primarily bacteria, have coevolved with humans over millions of years~\citep{rook_evolution_2017,martino_microbiota_2022,ma_systematic_2024}. The gastrointestinal tract harbors the most densely populated microbial community in the human body, supporting essential functions such as digestion, vitamin synthesis, and defense against pathogens~\citep{makki_impact_2018,pickard_gut_2017,canny_bacteria_2008,tarracchini_exploring_2024}. A balanced gut microbiota is necessary for maintaining these functions and long-term health~\citep{olvera-rosales_impact_2021}. Disruption of this balance leads to dysbiosis, an imbalance in the composition or function of the gut microbiota. Dysbiosis has been implicated in autoimmune, cardiometabolic, and psychiatric disorders~\citep{hashimoto_emerging_2023,valdes_role_2018,de_luca_microbiome_2019}. Several factors influence microbial equilibrium, including disease, antibiotic exposure, diet, age, lifestyle, environment, genetics, and physical activity~\citep{wen_factors_2017,weiss_normalization_2017,faust_conet_2016}. Therefore, each individual's gut microbiota is distinct and undergoes dynamic changes over time~\citep{rosenberg_diversity_2024}. The interactions among taxa continuously shape these temporal dynamic changes of microbial communities, motivating the study of microbial regulatory networks~\citep{wilkins_defining_2019,tiffany_dysbiosis_2019,krishnapriya_remodelling_2025,hrncir_gut_2022}.


Regulatory effects among taxa may be inhibitory, where one taxon suppresses the growth of another through mechanisms such as nutrient competition or production of inhibitory chemical compounds, or activatory, where one taxon enhances the growth of another by improving nutrient accessibility, producing beneficial metabolites, or generating new ecological niches~\citep{wang_microbial_2024,zhu_systematic_2025,canon_understanding_2020}. Network-based approaches have been developed to comprehensively capture the complex interdependencies among taxa~\citep{luo_progress_2024,champion_onenetone_2024}. A regulatory network consists of nodes and edges, where edges can be undirected, unidirectional, or omnidirectional. Undirected edges represent simple associations between nodes without implying direction of influence, while unidirectional edges indicate a one-way directional relationship from one node to another. Omnidirectional edges allow influence to propagate between a pair of nodes in two-way directions, though not necessarily with equal strength. Because observation of these interactions is rarely feasible, regulatory networks are typically inferred from abundance data measured at multiple discrete time points~\citep{guo_microbial_2022,momal_tree-based_2019}. However, abundance data are inherently noisy due to environmental fluctuations, sampling variations, and measurement errors, etc~\citep{lutz_survey_2022}. As a result, regulatory networks constructed from such data may contain inflated spurious edges, corresponding to interactions that lack biological meaning~\citep{faust_microbial_2012,lugo-martinez_dynamic_2019}. Thus, accurately constructing a regulatory network that reflects the underlying causal relationships among taxa from discretely observed temporal abundance data remains a significant challenge, further hindered by limitations in existing network construction methodologies~\citep{kehe_positive_2021,pacheco_multidimensional_2019}.

Numerous methods have been proposed for constructing regulatory networks, each offering distinct advantages and limitations. Boolean networks modeled each node as a binary variable, either 1 (active) or 0 (inactive), where Boolean logic defined the regulatory relationships governing state transitions between interacting nodes~\citep{rafimanzelat_global_2025}. However, Boolean networks were constrained by their inability to capture complex nonlinear relationships and by their neglect of dynamic trends~\citep{kadelka_meta-analysis_2024,lu_high-dimensional_2011}. Bayesian networks have been widely used to construct regulatory networks, in which edges represented conditional dependencies between nodes, and conditional distributions defined the probability of each node given the states of its parent nodes~\citep{puga_bayesian_2015,xing_improved_2017,sazal_inferring_2020,ruiz-perez_dynamic_2021}. Bayesian networks were unidirectional networks in which reverse edges (e.g., $X_1 \rightarrow X_2$ and $X_2 \rightarrow X_1$) were not permitted and additionally they were constrained by high computational costs~\citep{nagarajan_bayesian_2013,ghahramani_learning_1998,perrin_gene_2003,lu_high-dimensional_2011}. Mutual information quantified reduction in uncertainty about one variable given knowledge of another and was applied to regulatory network construction because it captured nonlinear dependencies~\citep{francis_comparative_2024,lei_approach_2023,butte_mutual_1999,dionisio_mutual_2004,song_comparison_2012,mokhtari_filtering_2022}. However, mutual information tended to overestimate direct interactions, as it could not distinguish between direct and indirect dependencies, leading to an inflated number of 
spurious edges~\citep{lei_approach_2023}. Furthermore, networks constructed using mutual information were undirected~\citep{chang_inference_2024}.

Ordinary Differential Equation (ODE)-based approaches have enabled the construction of omnidirectional regulatory networks from dynamic time-course data. Each node is represented by a time-dependent state variable describing its trajectory. Then the dynamics of these state variables are modeled by an ODE system, where each equation expresses the rate of change of one state variable as a function of all $p$ state variables. These methods are classified as parametric or nonparametric, depending on whether the functional form of the ODE system is specified prior to model fitting~\citep{sima_inference_2009,lu_high-dimensional_2011}. Parametric ODE models have typically assumed linear or sigmoidal (Hill-type) functions to represent regulatory effects~\citep{ehsan_elahi_method_2018}. These assumptions yielded models that were computationally efficient and straightforward to estimate from data, which gained popularity in constructing regulatory networks in biology~\citep{sima_inference_2009}. However, a fixed functional form may fail to capture the complex and heterogeneous nonlinear relationships encountered in practice~\citep{gerber_dynamic_2014,gonze_microbial_2018}. ODE-based nonparametric methods address this limitation by learning regulatory functions directly from data while preserving an underlying, but unknown, ODE structure. Existing nonparametric approaches typically assume an additive structure, expressing the regulatory function governing the rate of change of each state variable as a sum of $p$ univariate nonparametric functions, each corresponding to a potential regulator. This assumption serves two main purposes. First, it makes nonparametric estimation computationally feasible by decomposing a multivariate function into a collection of univariate functions. Second, it provides a direct interpretation of the network structure because each univariate function corresponds to a regulatory effect, thereby identifying an edge in the network~\citep{wu_sparse_2014, henderson_network_2014}.

These nonparametric methods can be further classified into derivative-based and integral-based approaches according to the ODE estimation procedure. Derivative-based methods first smoothed the discretely observed state trajectories, estimated their derivatives, and then fitted an additive nonparametric model that expressed each estimated derivative as a function of the $p$ smoothed state trajectories~\citep{wu_sparse_2014,henderson_network_2014}. Network Reconstruction via Dynamic Systems (NeRDS) was a derivative-based approach that estimated derivatives from smoothing splines and then fitted the additive nonparametric model using sparse backfitting~\citep{henderson_network_2014, ravikumar_sparse_2009}. The $L_2$ norm of each estimated univariate regulatory function was then utilized to quantify regulatory strength and rank all potential edges. However, NeRDS did not provide a principled criterion for determining which estimated edges should be retained in the network~\citep{henderson_network_2014}. Sparse Additive Ordinary Differential Equations (SA-ODE) was another derivative-based approach that incorporated an adaptive group LASSO penalty, identifying edges as those whose corresponding $L_2$ norms of the estimated regulatory functions were not shrunk to zero~\citep{wu_sparse_2014}.

\begin{figure}[tb]
\centering
\includegraphics[width=2.81in]{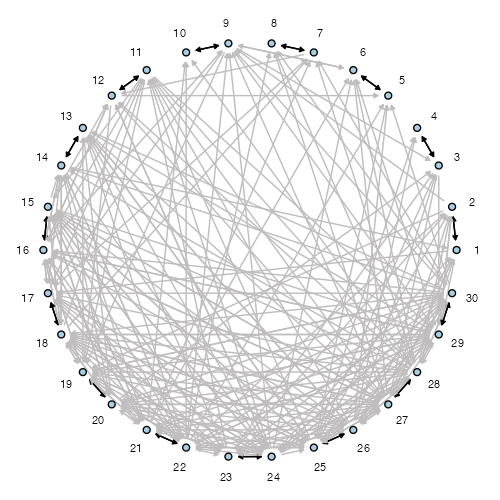}
\caption{Network constructed by the GRADE for Simulation~1-II (Section~\ref{sec:sim1-ib}). True edges are shown in black (\textcolor[HTML]{000000}{$\rightarrow$}) and spurious edges in gray (\textcolor[HTML]{BDBDBD}{$\rightarrow$}).
}
\label{fig:noisy}
\end{figure}

Integral-based approaches adopt the same additive nonparametric structure by modeling the smoothed trajectory of each state variable as a function of the $p$ integrated regulatory functions. Because the integral-based procedure depends only on the convergence of the estimated state trajectories, rather than on that of their derivatives, it achieves a faster convergence rate and has been shown to outperform derivative-based approaches both theoretically and empirically~\citep{chen_network_2017}. Graph Reconstruction via Additive Differential Equations (GRADE) was an integral-based approach that employed the group LASSO for edge screening~\citep{meier_group_2008,chen_network_2017}. Despite these advantages, the network constructed by GRADE often contained a large number of spurious edges because the group LASSO alone was insufficient to distinguish weak regulatory effects from estimation noise, as illustrated in Figure~\ref{fig:noisy}.

In this article, we propose a novel approach, Reconstruction of Enhanced Causal Omnidirectional Network (RECON), which leverages the integral-based additive nonparametric ODE structure to reconstruct regulatory networks from $p$ discretely observed time-course state trajectories. RECON contributes five methodological advances. First, it develops a new data-driven edge selection procedure by first quantifying the weighted strength of each regulatory function and then incorporating multiple components, including the group LASSO, Gaussian Mixture Model (GMM), and the maximum-ratio criterion, to adaptively determine an effective threshold for enhanced network reconstruction. This edge determination procedure substantially reduces the number of spurious edges while retaining the true regulatory edges, thereby greatly improving the accuracy of network reconstruction.
Second, RECON yields an omnidirectional regulatory network that captures causal relationships rather than merely statistical associations or noise artifacts, thereby improving both the reliability and interpretability of the reconstructed network. The reconstructed network also allows self-regulatory effects, in which a state variable influences its own dynamics. Third, existing ODE-based methods typically require dense and regular observations measured on a common time grid across all subjects. RECON greatly broadens the applicability of ODE-based network reconstruction by accommodating both dense regular and sparse irregular longitudinal data with flexible subject-specific sampling schedules and numbers of observations. Through an initial smoothing process, subject-specific trajectories are interpolated on a common dense time grid, allowing the standard ODE estimation procedure to be applied regardless of the original sampling scenario. Fourth, RECON enables a dynamic regulatory network in which both the trajectories of the state variable (node) and the regulatory effect functions (edges) evolve over time, with each regulatory effect modeled as a time-varying function rather than a single scalar coefficient. Compared with methods that estimate constant regulatory effects, such as the log-ratio penalized generalized estimating equation (FLORAL-GEE) method~\citep{fei_correlating_2025}, the estimated functional regulatory effects reveal how regulatory relationships strengthen, weaken, or reverse over time, truly reflecting the dynamic nature of real-world biological systems. Fifth, RECON outputs the signs and weights of regulatory edges, enabling comprehensive interpretation of network architecture. The sign of each regulatory edge is determined by integrating the estimated regulatory function over its entire domain, allowing each edge to be classified as either activatory (defined by a positive sign) or inhibitory (defined by a negative sign). The weighted strength of each regulatory edge is quantified by the norm of the coefficients of the corresponding estimated regulatory function. In addition, RECON identifies keystone nodes through degree and betweenness centralities and characterizes the topological organization of the network.

We evaluate the performance of RECON through five simulation studies spanning a wide range of network structures, network sizes, and levels of complexity. Across all settings, RECON consistently outperforms GRADE, which relies exclusively on the group LASSO. In all five simulation studies, RECON reduces the number of spurious edges to zero or near zero while retaining nearly all true regulatory edges, demonstrating high reconstruction accuracy, even under a non-additive ODE system. In the most challenging scenario, RECON reduces the number of spurious edges from 239 to 0.

The real data analysis focuses on the longitudinal gut microbiota dataset collected from allogeneic hematopoietic cell transplantation (allo-HCT) patients at Memorial Sloan Kettering Cancer Center (MSK)~\citep{liao_compilation_2021}. This same dataset has previously been analyzed from different perspectives to address a variety of scientific questions using substantially different analytical strategies with distinct research foci~\citep{yan_compilation_2022,schluter_taxumap_2023,nguyen_high-resolution_2023,liao_oral_2024,fei_scalable_2024,stamper_perturbation-aware_2025,mcgovern_scale_2026,qureshi_dynabiome_2026}. For example, \citet{fei_scalable_2024} proposed FLORAL, which fitted a log-ratio LASSO regression model to identify bacterial genera associated with clinical outcomes, such as post-transplant survival. For each patient, FLORAL related a scalar outcome to a weighted sum of the log-ratios of all $\binom{p}{2}$ pairs of genera abundances across all repeatedly measured samples, where $p=220$ genera. A LASSO penalty was then applied and several bacterial genera associated with patient survival outcomes were identified. However, FLORAL treated the repeatedly measured abundance values as independent predictor observations and did not explicitly model their temporal dynamics. \citet{stamper_perturbation-aware_2025} proposed a perturbation-aware Neural ODE (pNODE) to model the dynamic trajectories and predict future bacterial abundances and bloodstream infections. Their analysis was conducted at the class level, using $p=13$ bacterial classes. The multilayer perceptron (MLP) underlying pNODE captured complex nonlinear dynamics well but offered limited insight into detailed regulatory relationships, as its black-box hidden-layer representations could not be directly interpreted.

To the best of our knowledge, RECON is the first study applied to this MSK hospitalome dataset to reconstruct a causal omnidirectional network that comprehensively characterizes interactions among bacterial taxa. Beyond identifying causal regulatory relationships, RECON reconstructs the complete network topology and quantifies the two-way direction, sign, and strength of every regulatory edge. The resulting pre- and post-HCT networks exhibit substantially different topological structures and taxonomic compositions, providing new biological insights into microbiota regulation during transplantation. Furthermore, the estimated regulatory trajectories characterize how the influence of each taxon on others evolves over time, revealing dynamic regulatory patterns that cannot be inferred from abundance measurements alone.

\section{Data Background}\label{sec:data}

The MSK hospitalome is a comprehensive large-scale dataset compiled from leukemia and lymphoma patients undergoing allo-HCT~\citep{liao_compilation_2021}. During allo-HCT, each patient's immune system was first ablated through a conditioning regimen and subsequently reconstituted via infusion of donor stem cells, resulting in severe disruption of the gut microbiota, undergoing a period of near-complete microbial depletion followed by gradual recovery. The hospitalome comprises 12,546 stool samples collected from 1,278 patients with sparse and irregular sampling before and after transplantation. Two non-overlapping time windows were defined relative to day~0, the day of transplantation: a pre-HCT window spanning days $[-15,0)$ and a post-HCT window spanning days $[0,35]$. This dataset overcomes limitations common to many human microbiota studies, including small sample sizes, cross-sectional designs, and insufficient clinical metadata to account for confounding factors~\citep{liao_compilation_2021}.

Our analysis focuses on the repeatedly measured time-course abundance data for $29$ bacterial families identified across the pre- and post-HCT time windows, where \emph{family} is a standard taxonomic rank. Some families are observed in both time windows and others only observed in either pre-HCT window or the post-HCT window. The repeated sampling design, offering a sufficient number of observations per patient, and large patient cohort make this dataset particularly well suited for the proposed RECON approach.

\begin{figure}[tb]
\centering

\begin{subfigure}{5.62in}

\includegraphics[width=\linewidth]{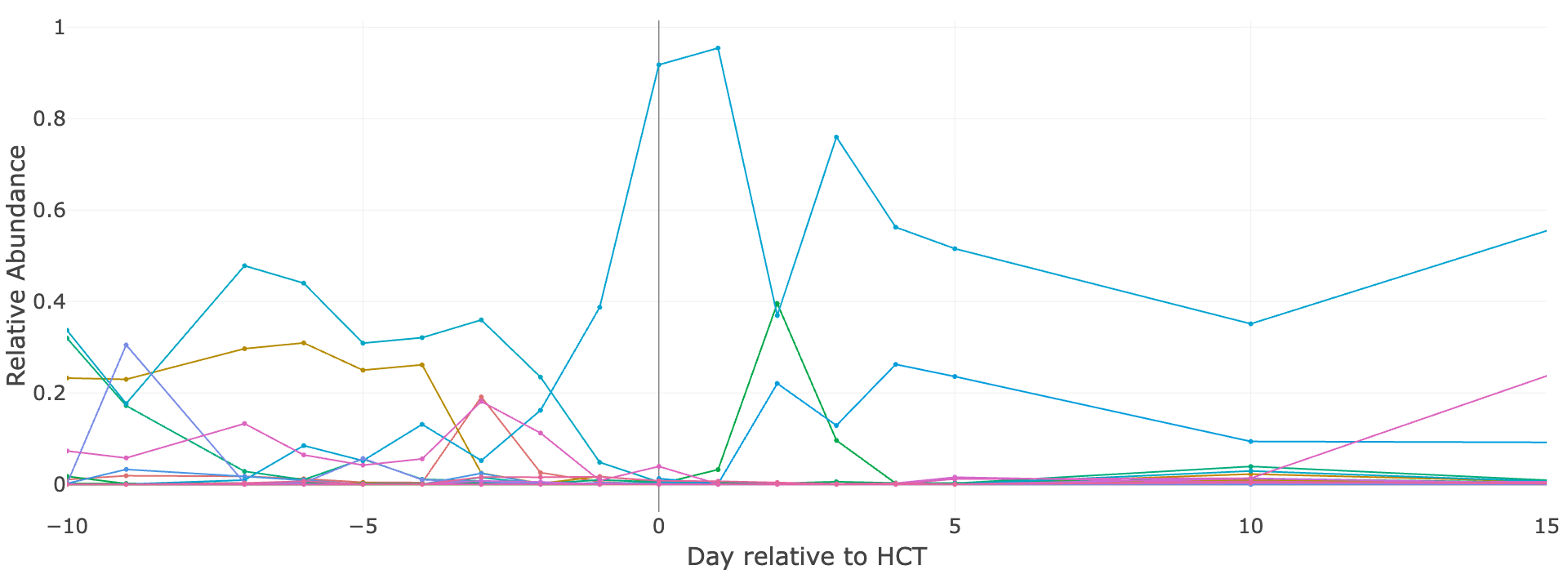}
\caption{}
\label{fig:onepallb-1}
\end{subfigure}

\begin{subfigure}{5.62in}

\includegraphics[width=\linewidth]{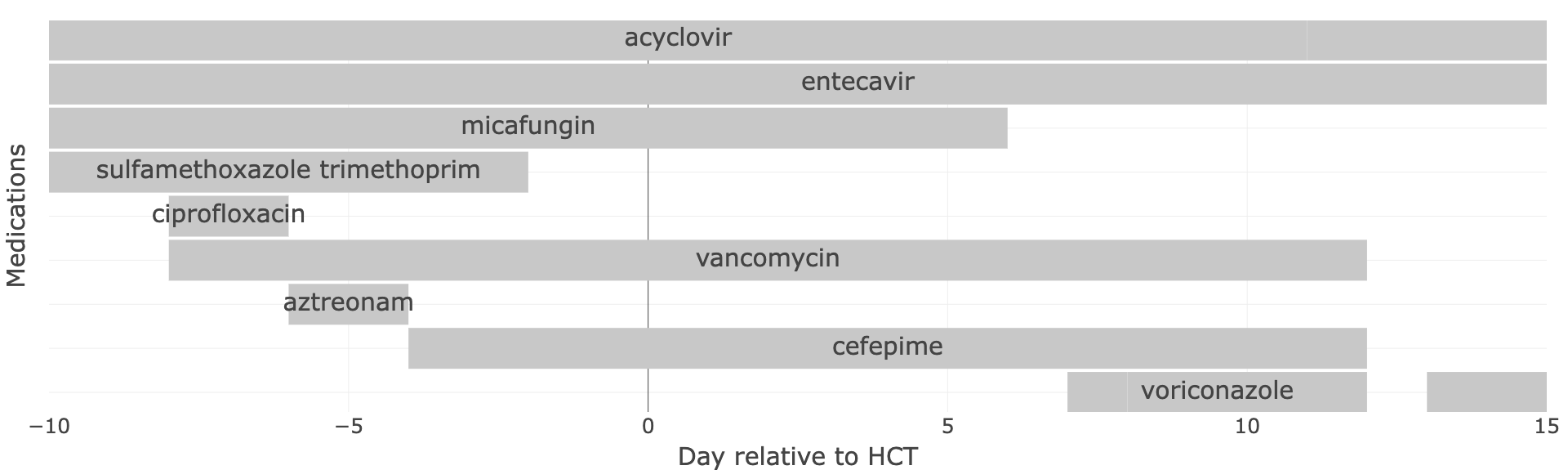}
\caption{}
\label{fig:onepallb-2}
\end{subfigure}

\caption{All $29$ bacterial families and medication exposure for one representative patient (PatientID~1207), shown for days $-10$ to $+15$ relative to transplantation (Day~0). Panel (a): The discretely measured relative abundance values of all bacterial families identified across the pre- and post-HCT study periods, recorded at $n=16$ discrete sampling time points. Panel (b): Medication exposure timeline, where each row corresponds to one medication and the gray bars indicate the duration of administration.}
\label{fig:onepallb}
\end{figure}

Figure~\ref{fig:onepallb} illustrates the observed microbiota data for a single patient (PatientID~1207). Panel~\ref{fig:onepallb-1} presents the relative abundances of $29$ bacterial families measured repeatedly at discrete time points spanning days $-10$ to $+15$ relative to transplantation, where day~0 denotes the day of transplantation. Most bacterial families exhibit zero or near-zero relative abundance, whereas only a small subset exhibits substantial variation in both magnitudes and temporal trajectories throughout the study period. Panel~\ref{fig:onepallb-2} depicts the patient's medication exposure over the study period. This patient received multiple medications, with some administered throughout the entire study period and others only during brief time intervals.

\begin{figure}[tb]
\centering

\begin{subfigure}{5.62in}

\includegraphics[width=\linewidth]{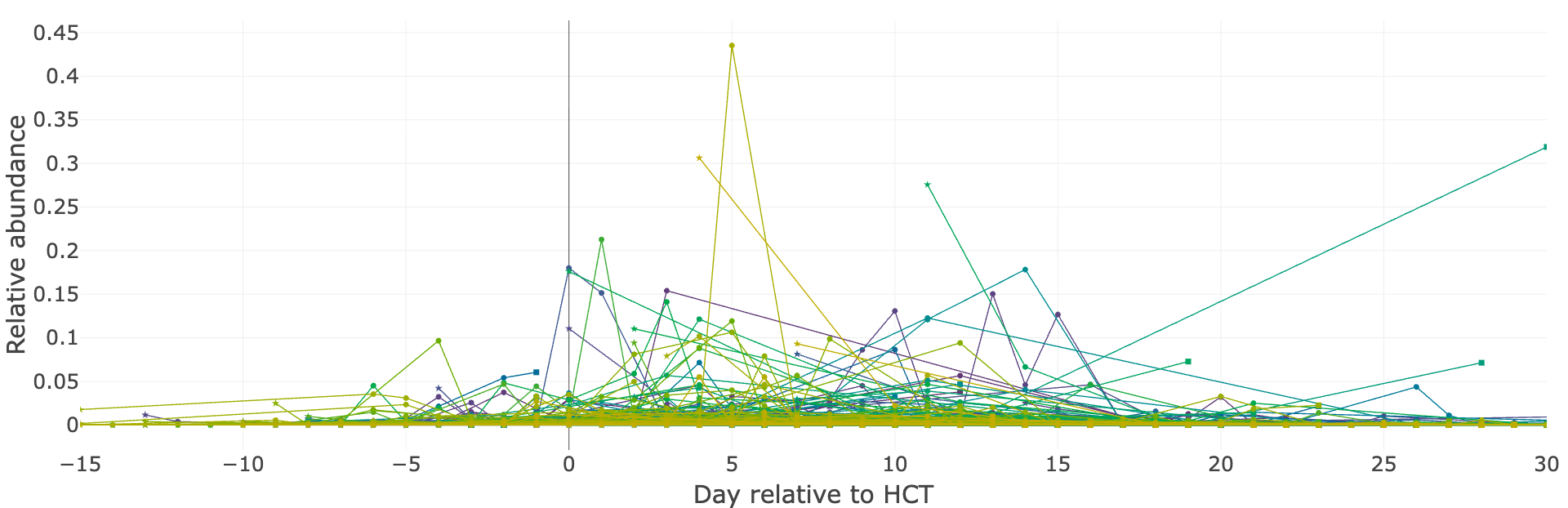}
\caption{}
\label{fig:allponeb-1}
\end{subfigure}

\begin{subfigure}{5.62in}

\includegraphics[width=\linewidth]{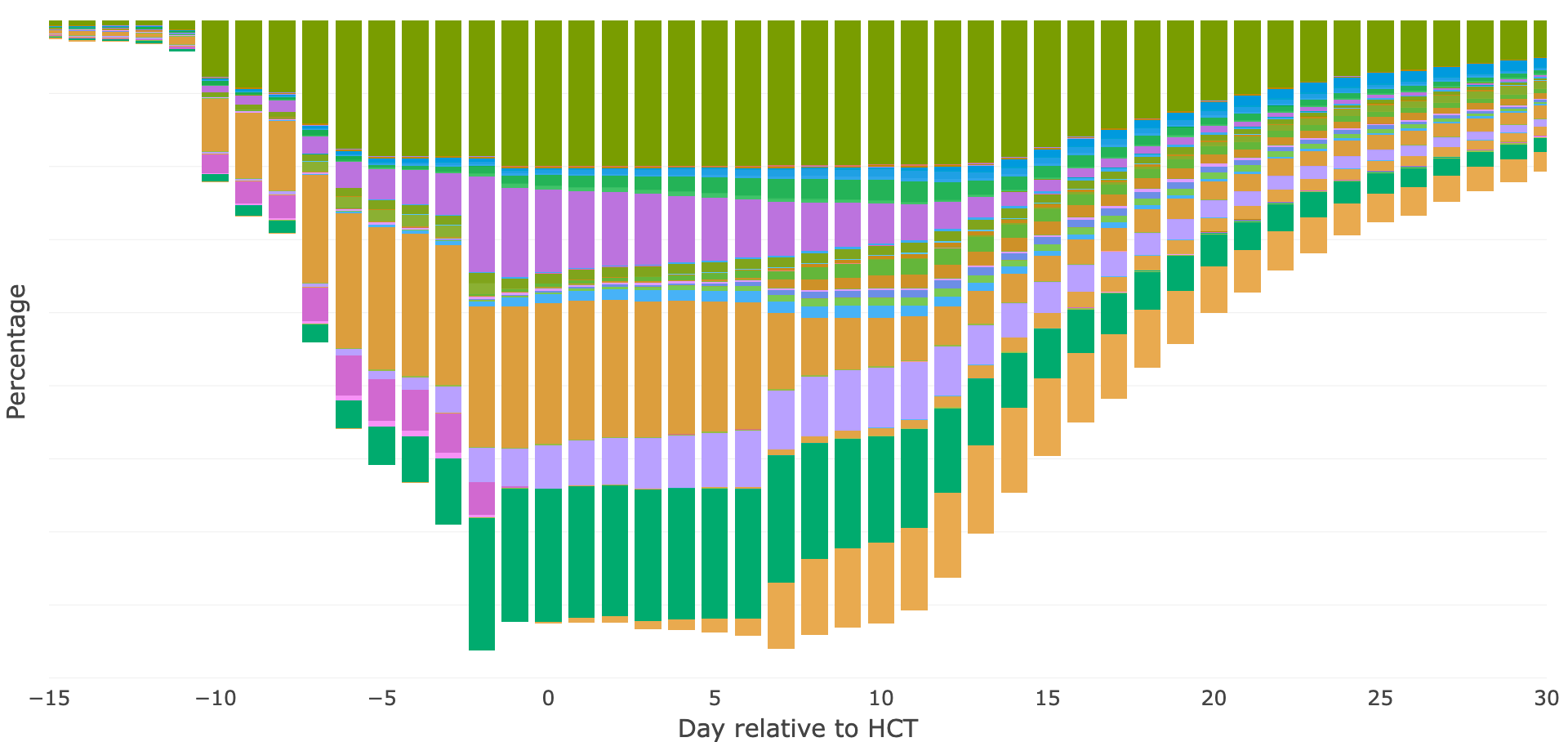}
\caption{}
\label{fig:allponeb-2}
\end{subfigure}
\caption{Relative abundance data of one bacterial family (\textit{Family XI}) for all patients, together with their medication exposure over time. Day~$0$ denotes the transplantation day, and both panels span days $-15$ to $+30$. Panel (a): The discretely observed relative abundance values for the $741$ patients, measured at subject-specific sampling schedules with varying numbers of observation time points, where $\star$ and $\square$ denote the first and last recorded samples for each patient, respectively. Panel (b): Percentage of the same $741$ patients receiving each medication at each time point. Colors correspond to the medication names shown in Appendix Figure S1.}
\label{fig:allbonep}
\end{figure}

Figure~\ref{fig:allbonep} illustrates the observed microbiota data for a single bacterial family (\textit{Family~XI}). Panel~\ref{fig:allponeb-1} presents the relative abundances of \textit{Family~XI} measured repeatedly for $741$ patients at discrete time points spanning days $-15$ to $+30$ relative to transplantation, where day~$0$ denotes the day of transplantation. Even for the same bacterial family, relative abundances vary substantially across patients and over time, exhibiting highly heterogeneous temporal trajectories. Many patients remain at near-zero relative abundance, whereas others exhibit substantial temporal fluctuations. Patients were sampled either before transplantation only, after transplantation only, or during both study periods. Panel~\ref{fig:allponeb-2} illustrates the percentage of patients receiving each medication at each observed time point. As shown, patients received the greatest variety of medications around the time of transplantation, approximately from days $-5$ to $+10$, with the variety of medications gradually decreasing toward both ends of the study period.

In summary of Figures~\ref{fig:onepallb} and~\ref{fig:allbonep}, we emphasize an important property of the data structure. For each patient, all bacterial families are observed at exactly the same sampling time points. However, these shared sampling time points differ from one patient to another, resulting in an irregular and sparsely sampled longitudinal dataset.

\section{Methodology}\label{sec:method}


\subsection{Model Formulation and Initial Smoothing}\label{sec:formulation}
We now describe the statistical model, starting at the level of a single subject. Let $p$ be the number of state variables and $n$ be the number of observed time points. Let $\mathbf{Y}_{i}=(Y_{i1},\ldots,Y_{ip})^\top\in\mathbb{R}^p$ represent the original noisy observations of the $p$ state variables measured at the discrete time point $t_i\in[0,1]$, where time has been rescaled from the original study period to the unit interval, so that
\begin{equation}\label{eq:obs}
\mathbf{Y}_i
=
\mathbf{X}(t_i;\boldsymbol{\theta})
+
\boldsymbol{\varepsilon}_i,
\quad
i=1,\ldots,n,
\end{equation}
where $\boldsymbol{\varepsilon}_i = (\varepsilon_{i1}, \ldots, \varepsilon_{ip})^{\top} \in \mathbb{R}^p$ represents measurement errors. We assume that $\varepsilon_{ij} 
\overset{\text{i.i.d.}}{\sim} \mathscr{N}(0,\sigma^2)$ for $i = 1,\ldots,n$ and $j = 1,\ldots,p$. Let 
$\mathbf{X}(t;\boldsymbol{\theta})=
(X_1(t;\boldsymbol{\theta}),\ldots,X_p(t;\boldsymbol{\theta}))^\top$ denote the underlying $p$-dimensional smooth state trajectory as a function of time $t$, parameterized by $\boldsymbol{\theta}$. Its $j$th component, $X_j(t;\boldsymbol{\theta})$, represents the smooth trajectory of state variable $X_j$, $j=1,\ldots,p$.

The rate of change of $\mathbf{X}(t;\boldsymbol{\theta})$ is governed by the ODE system
\begin{equation}\label{eq:ode}
\mathbf{X}'(t;\boldsymbol{\theta})
=
\mathbf{f}\big(\mathbf{X}(t;\boldsymbol{\theta}),\boldsymbol{\theta}\big),
\quad t\in[0,1],
\end{equation}
where $\mathbf{f}=(f_1,\ldots,f_p)^\top$ is a $p$-dimensional function that is twice continuously differentiable with bounded $L_2$ norm. Since the functional form of $\mathbf{f}$ is unknown, we adopt an additive nonparametric approach for its estimation. Accurate estimation of $\mathbf{X}(t;\boldsymbol{\theta})$, $\mathbf{f}$ and the parameter $\boldsymbol{\theta}$ in the ODE system in Equation~\eqref{eq:ode} is critical for learning the complex dynamics of regulatory effects among the state variables and hence achieving accurate network constructions.

Since the functions $X_1(t;\boldsymbol{\theta}), \ldots,
X_p(t;\boldsymbol{\theta})$ are unknown, we first obtain their initial smooth estimates, denoted by
$\widetilde{X}_1(t;\boldsymbol{h}), \ldots,
\widetilde{X}_p(t;\boldsymbol{h})$.
These smoothed trajectories are then used to estimate the unknown function $\mathbf{f}$ in Equation~\eqref{eq:ode}, and the resulting estimate $\widehat{\mathbf{f}}$ is used to obtain the final trajectory estimates $\widehat{X}_1(t;\mathbf{h},\boldsymbol{\theta}), \ldots, \widehat{X}_p(t;\mathbf{h},\boldsymbol{\theta})$.

Equation~\eqref{eq:obs} describes the model formulation for a single subject. We now extend the same setup to multiple subjects, assuming that all $p$ state variables are observed at the same common and dense time points across all subjects. Let $R$ denote the number of subjects, with larger $R$ generally improving the accuracy of parameter estimation~\citep{wu_sparse_2014,lu_high-dimensional_2011}. Then, for each subject $r=1,\ldots,R$, Equation~\eqref{eq:obs} becomes
\begin{equation}\label{eq:obs-rep}
\mathbf{Y}^{(r)}_i
=
\mathbf{X}^{(r)}(t_i;\boldsymbol{\theta})
+
\boldsymbol{\varepsilon}^{(r)}_i,
\qquad
i=1,\ldots,n.
\end{equation}
We assume that the trajectories of all subjects are governed by the same ODE system in Equation~\eqref{eq:ode} with common parameter $\boldsymbol{\theta}$ but different initial values, thereby yielding subject-specific trajectories.

Standard ODE modeling generally requires that all subjects be observed at a common set of sufficiently dense sampling time points for all state variables, as in Equation~\eqref{eq:obs-rep} does. However, these assumptions are often violated in practice, particularly in biomedical longitudinal studies. To make the proposed methodology applicable to a wider range of settings, we consider two sampling scenarios separately. In both scenarios, the initial state trajectories are obtained via nonparametric smoothing and subsequently interpolated onto a dense and common time grid shared by all subjects and all state variables. Different smoothing strategies are employed according to the sampling scenarios: (1) dense and regular observations and, (2) sparse and irregular observations.

\subsubsection{Local polynomial smooth for the dense and regular scenario} \label{sec:method-locpoly}
If the observed sampling time points are dense and regular, that is, all subjects are observed on a common set of time points $t_i$, $i=1,\ldots,n$, as in Equation~\eqref{eq:obs-rep}, then we apply local polynomial smoothing to obtain the initial smoothed state trajectories~\citep{fan_local_2018,loader_local_1999}. Specifically, for each subject $r=1,\ldots,R$ and each state variable $j=1,\ldots,p$, the smoothed trajectory is obtained by
\begin{equation}\label{eq:locpoly-r}
{\widetilde{X}_j}^{(r)}(t;\mathbf{h})
=
\arg\min_{Z_j(t)\in\mathcal{X}(\mathbf{h})}
\sum_{i=1}^n
\left(
Y^{(r)}_{ij}
-
Z_j(t_i)
\right)^2,
\end{equation}
where the class $\mathcal{X}(\mathbf{h})$ consists of all candidate functions, $Z_j(t)\in\mathcal{X}(\mathbf{h})$, obtained from local polynomial regression at a chosen polynomial degree with candidate bandwidth $\mathbf{h}$ selected using generalized cross-validation (GCV). Note that local polynomial smoothing is performed separately for each state variable $X_j$ and each subject $r$. Consequently, the sum term in Equation~\eqref{eq:locpoly-r} only involves time points without borrowing information across subjects during the smoothing process.

\subsubsection{PACE smooth for the sparse and irregular scenario}\label{sec:pace}

If the observed time points are sparse and irregular across subjects, then we employ the Principal Analysis by Conditional Expectation (PACE) method to obtain the initial smooth estimates of $\mathbf{X}(t;\boldsymbol{\theta})$. Unlike local polynomial smoothing, PACE borrows strength across all subjects and obtains subject-specific trajectories through functional principal component scores estimated via conditional expectation~\citep{yao_functional_2005,zhou_fdapace_2016}.

To accommodate this setting, we introduce a more general notation. Let $t_{r1}, \ldots, t_{rn_r}$ denote the discretely observed sampling time points and $n_r$ be the number of time points for subject $r=1, \ldots, R$. Let $Y^{(r)}_{ij}$ denote the original noisy observation of the $j$th state variable at time point $t_{ri}$ for subject $r$. Then,
\begin{equation}\label{eq:obs-pace}
Y^{(r)}_{ij}
=
X^{(r)}_j(t_{ri};\boldsymbol{\theta})
+
\varepsilon^{(r)}_{ij},
\qquad
 r=1,\ldots, R, \quad i = 1, \ldots, n_r, \quad j=1,\ldots,p,
\end{equation}
where $\varepsilon^{(r)}_{ij}$ are independent mean-zero random errors with variance $\sigma_j^2$.

For each $j = 1, \ldots, p$, the mean function $\mu_j(t) = E[X^{(r)}_j(t;\boldsymbol{\theta})]$ is estimated by local linear smoothing of the pooled observations $\{(t_{ri}, Y^{(r)}_{ij})\}$ across all subjects, yielding $\widehat{\mu}_j(t)$. The estimated covariance surface, $\widehat{\mathcal{D}}_j(s,t)$, is obtained by applying local linear surface smoothing to the following pairwise products
\begin{equation*}
\bigl(Y^{(r)}_{ij} - \widehat{\mu}_j(t_{ri})\bigr)
\bigl(Y^{(r)}_{lj} - \widehat{\mu}_j(t_{rl})\bigr),
\qquad i \neq l, \quad i, l \in \{1, \ldots, n_r\},\quad r=1,\ldots,R,
\end{equation*}
pooled across all subjects~\citep{yao_functional_2005}. The bandwidths $\mathbf{h}$ for the mean function and covariance surface are selected via GCV. Then the eigen-decomposition of $\widehat{\mathcal{D}}_j(s,t)$ yields estimated eigenfunctions $\widehat{\phi}_{j1}(t),\widehat{\phi}_{j2}(t),\ldots$, and the corresponding non-increasing eigenvalues $\widehat{\Lambda}_{j1} \geq \widehat{\Lambda}_{j2} \geq \cdots \geq 0$. 

Denote $\mathbf{Y}^{(r)}_{\cdot, \, j} = (Y^{(r)}_{1j}, \ldots, Y^{(r)}_{n_r j})$ and $\widehat{\boldsymbol{\mu}}^{(r)}_j = (\widehat{\mu}_j(t_{r1}), \ldots, \widehat{\mu}_j(t_{rn_r}))$, the functional principal component scores for subject $r$ are estimated by conditional expectation as
\begin{equation}\label{eq:pace-score}
\widehat{\xi}^{(r)}_{jv}
= \widehat{\Lambda}_{jv}\,
\widehat{\boldsymbol{\phi}}^{(r)}_{jv}\,
\left(\widehat{\boldsymbol{\Sigma}}^{(r)}_j\right)^{-1}
\big((\mathbf{Y}^{(r)}_{\cdot, \, j})^\top - (\widehat{\boldsymbol{\mu}}^{(r)}_j)^\top\big),~~v=1,\ldots,A_j,
\end{equation}
where $\widehat{\boldsymbol{\phi}}^{(r)}_{jv} = (\widehat{\phi}_{jv}(t_{r1}), \ldots, \widehat{\phi}_{jv}(t_{rn_r}))$ and the $(i,l)$-th entry of $\widehat{\boldsymbol{\Sigma}}^{(r)}_j$ is $\widehat{\mathcal{D}}_j(t_{ri}, t_{rl}) + \widehat{\sigma}^2_j\, \mathbbm{1}_{\{i=l\}}$. The number of retained eigenfunctions, $A_j$, is selected separately for each $j$ by the Akaike information criterion (AIC)~\citep{yao_functional_2005}. Then the trajectory for state variable $j$ and subject $r$ is estimated as
\begin{equation}\label{eq:pace-r}
\widetilde{X}_j^{(r)}(t;\mathbf{h})
= \widehat{\mu}_j(t)
+ \sum_{v=1}^{A_j} \widehat{\xi}^{(r)}_{jv}\,\widehat{\phi}_{jv}(t).
\end{equation}
The bandwidth $\mathbf{h}$ is selected by GCV. Once the smooth trajectory $\widetilde{X}_j^{(r)}(t;\mathbf{h})$ is obtained for each subject $r$, it is interpolated onto a dense and common time grid $t_1,\ldots,t_n$ shared by all subjects. Consequently, the original irregular and sparse observations in Equation~(\ref{eq:obs-pace})  are transformed into a common, regular, and densely sampled representation, as described in Equation~(\ref{eq:obs-rep}), which is required for standard ODE estimation. Note that the same procedure from Equation (\ref{eq:obs-pace}) to Equation (\ref{eq:pace-r}) is performed independently and separately for each state trajectory, $j=1,\ldots,p$. 

\subsection{Additive Nonparametric ODE Structure}\label{sec:method-additive}
We will now setup the estimation of the regulatory functions $\mathbf{f}$ in Equation~\eqref{eq:ode}. Since the true form of $\mathbf{f}$ is unknown, a nonparametric estimation is adopted without assuming a pre-specified functional form for it~\citep{eubank_nonparametric_1999,henderson_network_2014}. To make nonparametric estimation of $\mathbf{f}$ tractable, we assume the additive structure where each component $f_j$ of $\mathbf{f}$ is a sum of $p$ univariate functions $f_{jk}$, which depend solely on $X_k$ as follows \citep{henderson_network_2014, wu_sparse_2014}:
\begin{equation}\label{eq:additive}
f_j(\mathbf{X}(t;\boldsymbol{\theta})) = \sum_{k=1}^{p} f_{jk} (X_k (t; \boldsymbol{\theta})),
\end{equation}
The combination of Equations~\eqref{eq:ode} and~\eqref{eq:additive} yields an additive nonparametric ODE system, 
\begin{equation}\label{eq:additive-ode}
X_j'(t;\boldsymbol{\theta})=\theta_{j0}+\sum_{k=1}^{p}f_{jk}\!\big(X_k(t;\boldsymbol{\theta}) \big), \quad t\in[0,1],\quad \theta_{j0}\in \mathbb{R}, \quad j = 1, \ldots, p.
\end{equation}
where $f_{jk}(X_k(t;\boldsymbol{\theta}))$ 
directly quantifies the regulatory effect of state variable $X_k$ on 
state variable $X_j$, for each $k = 1, \ldots, p$~\citep{wu_sparse_2014,henderson_network_2014,chen_network_2017}. 

As a nonparametric smooth function, $f_{jk}(\cdot)$ can be estimated using a finite number of basis functions, for example a B-spline basis $\boldsymbol{\psi}(\cdot)=(\psi_1(\cdot),\ldots,\psi_M(\cdot))^\top$, so that
\begin{equation}\label{eq:basis}
f_{jk}(\cdot)=\boldsymbol{\psi}(\cdot)^\top\boldsymbol{\theta}_{jk}+\delta_{jk}(\cdot),\qquad \boldsymbol{\theta}_{jk}\in\mathbb{R}^M. 
\end{equation}
Here $M$ denotes the number of basis functions, 
$\delta_{jk}(\cdot)$ is the residual function that is assumed 
to be small in practice~\citep{chen_network_2017}. The additive structure in Equation~\eqref{eq:additive-ode} together with the basis expansion in Equation~\eqref{eq:basis} now render $\boldsymbol{\theta}$ as a $p\times p\times M$ array with $\boldsymbol{\theta}[j,k,\cdot] = \boldsymbol{\theta}_{jk} = (\theta_{jk,1},\ldots,\theta_{jk,M})^\top \in \mathbb{R}^M$ for $j,k=1,\ldots,p$. Substituting Equation~\eqref{eq:basis} into the additive ODE system in Equation~\eqref{eq:additive-ode} yields:
\begin{equation}\label{eq:additive-ode-basis}
\begin{split}
X'_j(t;\boldsymbol{\theta})=\theta_{j0}+\sum_{k=1}^p \boldsymbol{\psi}\!\big(X_k(t;\boldsymbol{\theta})\big)^\top \boldsymbol{\theta}_{jk}
       + \sum_{k=1}^p \delta_{jk}\!\big(X_k(t;\boldsymbol{\theta})\big),\\ 
\quad t\in[0,1], \quad \theta_{j0}\in \mathbb{R}, \quad j=1,\ldots,p. 
\end{split}
\end{equation}
Integrating Equation~\eqref{eq:additive-ode-basis} over the interval $[0,t]$ yields
\begin{equation}\label{eq:integrated}
\begin{split}
X_j(t;\boldsymbol{\theta}) = X_j(0) + \theta_{j0}\,t 
+ \sum_{k=1}^p \boldsymbol{\Psi}_k(t;\boldsymbol{\theta})^\top\boldsymbol{\theta}_{jk}
+ \sum_{k=1}^p \int_0^t \delta_{jk}\!\big(X_k(u;\boldsymbol{\theta})\big)\,du,\\
\quad t\in[0,1], \quad \theta_{j0}\in \mathbb{R}, \quad j=1,\ldots,p,
\end{split}
\end{equation}
where
\begin{equation}\label{eq:int-basis}
\boldsymbol{\Psi}_k(t;\boldsymbol{\theta})=(\Psi_{k1}(t;\boldsymbol{\theta}),\ldots ,\Psi_{kM}(t;\boldsymbol{\theta}))^\top
=\!\int_0^t \boldsymbol{\psi}\!\big(X_k(u;\boldsymbol{\theta})\big)\,du
\end{equation}
are the integrated basis functions.

\subsection{Estimating the Unknown Parameters}\label{sec:method-theta}
 Some modifications are made when moving from the population-level formulation in Equation~\eqref{eq:integrated} to the estimation process. The residual integral term $\displaystyle\sum_{k=1}^p\displaystyle\int_0^t\delta_{jk}(X_k(u;\boldsymbol{\theta}))\,du$ is treated as negligible and hence dropped. Moreover, since the true initial value $X_j(0)$ is unobserved, it is replaced by a subject-specific intercept parameter $C_{j0}^{(r)}\in\mathbb{R}$, estimated jointly with $\boldsymbol{\theta}_{jk}$. 
 
 Now, the state trajectory for each subject $r$ and each $j$ is estimated as
\begin{equation}\label{eq:tildeX}
\widehat{X}^{(r)}_j(t;\mathbf{h},\boldsymbol{\theta})
= \widehat{C}^{(r)}_{j0}+\widehat{\theta}_{j0}t
+\sum_{k=1}^p \widehat{\boldsymbol{\Psi}}^{(r)}_k(t;\mathbf{h})^\top\boldsymbol{\theta}_{jk},
\end{equation}
where  
\begin{equation}\label{eq:Psi-rep}
\widehat{\boldsymbol{\Psi}}^{(r)}_k(t;\mathbf{h})=\displaystyle\int_0^t \boldsymbol{\psi}\!\big(\widetilde{X}^{(r)}_k(u;\mathbf{h})\big)\,du, \quad k=1,\ldots,p.
\end{equation}
Note that we substitute the initial smooth $\widetilde{X}^{(r)}_j(t;\mathbf{h})$, obtained from Equations~\eqref{eq:locpoly-r} or~\eqref{eq:pace-r}, into Equation~\eqref{eq:Psi-rep} because 
$X_k(t;\boldsymbol{\theta})$ is unknown. But $\widetilde{X}^{(r)}_j(t;\mathbf{h})$ is merely an initial smooth estimate that depends solely on the observed data and does not necessarily satisfy the ODE system in Equation~\eqref{eq:additive-ode}. In contrast, $\widehat{X}^{(r)}_j(t;\mathbf{h},\boldsymbol{\theta})$, estimated from Equation~\eqref{eq:tildeX}, satisfies the ODE structure in Equation~\eqref{eq:additive-ode}~\citep{chen_network_2017}.

The estimate of unknown parameter $\boldsymbol{\theta}$ is obtained by solving the following optimization problem for each $j=1,\ldots,p$,
\begin{equation}\label{eq:opt-rep}
\begin{aligned}
\widehat{\boldsymbol{\theta}}_j = \arg\min_{\substack{C_{j0}^{(r)}\in\mathbb{R},\,\theta_{j0}\in\mathbb{R},\\ \boldsymbol{\theta}_{j1},\ldots,\boldsymbol{\theta}_{jp}\in\mathbb{R}^M}}
&\;\frac{1}{2Rn}\sum_{r=1}^{R}\sum_{i=1}^{n}
\Big\{Y^{(r)}_{ij}-\widehat{X}^{(r)}_j(t_i;\mathbf{h},\boldsymbol{\theta})\Big\}^2 \\
&\;+\;\lambda_{n,j}\sum_{k=1}^{p}\Bigg(\frac{1}{Rn}\sum_{r=1}^{R}\sum_{i=1}^{n}\{\widehat{\boldsymbol{\Psi}}_k^{(r)}(t_i;\mathbf{h})^\top\boldsymbol{\theta}_{jk}\}^2\Bigg)^{1/2},
\end{aligned}
\end{equation}
where $\widehat{\boldsymbol{\Psi}}_k^{(r)}(t;\mathbf{h})$ is defined as in Equation~\eqref{eq:Psi-rep}.

For each $j = 1,\ldots,p$, the parameter $\widehat{\boldsymbol{\theta}}_j = \{\widehat{\boldsymbol{\theta}}_{j1},\ldots,\widehat{\boldsymbol{\theta}}_{jp}\}$ has $p$ components, and each of its components $\widehat{\boldsymbol{\theta}}_{jk} = (\widehat{\theta}_{jk,1},\ldots,\widehat{\theta}_{jk,M})^\top$ has $M$ terms because of the basis coefficients setup from Equation~\eqref{eq:basis}. The group LASSO promotes sparsity at the group level by penalizing all $M$ coefficients in each component $\widehat{\boldsymbol{\theta}}_{jk}$ as a whole unit. The tuning parameter $\lambda_{n,j}$ is selected using the BIC~\citep{meier_group_2008}. 

\subsection{Regulatory Network Reconstruction}\label{sec:method-netrec}
In the regulatory network, each state variable $X_j$ is represented by node $j$, for $j=1,\ldots,p$. From a population perspective, a directed edge $k\rightarrow j$ exists if and only if $X_k$ regulates $X_j$, that is, if and only if $f_{jk}\not\equiv0$ in the ODE system in Equation~\eqref{eq:additive-ode}. By Equation~\eqref{eq:iff}, this is equivalent to $\|\boldsymbol{\theta}_{jk}\|_2\neq0$. From an estimation perspective, however, determining whether an edge should be included in the regulatory network is substantially more challenging. After obtaining the parameter estimates $\widehat{\boldsymbol{\theta}}_{jk}$ from Equation~\eqref{eq:opt-rep}, the strength of the regulatory edge, $k \rightarrow j$, is quantified by the magnitude of $\|\widehat{\boldsymbol{\theta}}_{jk}\|_2$. In practice, however, retaining all estimated edges with $\|\widehat{\boldsymbol{\theta}}_{jk}\|_2>0$ would result in a large number of spurious edges due to estimation error. Therefore, a threshold must be imposed on $\|\widehat{\boldsymbol{\theta}}_{jk}\|_2$ to determine the existence of the corresponding edge. Choosing an appropriate threshold is not a trivial problem and is critical to the accuracy of network reconstruction.

Let $G^*$ denote the population binary adjacency matrix of the true but unknown underlying regulatory network among the $p$ nodes, where $G^*[k,j] = 1$ if the true edge $k \rightarrow j$ exists and $G^*[k,j] = 0$ otherwise. The objective of network reconstruction is to obtain an estimated adjacency matrix  $\widehat{G}$ to approximate $G^*$ as accurately as possible.

Since
\begin{equation}\label{eq:iff}
f_{jk} \equiv 0 \quad \text{if and only if} \quad 
\|\boldsymbol{\theta}_{jk}\|_2 = 0,
\end{equation}
using the non-zero estimates $\widehat{\boldsymbol{\theta}}_{jk}$ obtained from the group LASSO in Equation~\eqref{eq:opt-rep}, the network can be constructed from the adjacency matrix $\widehat{G}_{\text{GRADE}}$ defined as follows: 
\begin{equation}\label{eq:g-bic}
\widehat{G}_{\text{GRADE}}[k,j] = 
{\mathbbm{1}}_{\{\|\widehat{\boldsymbol{\theta}}_{jk}\|_{2}
> 0\}},~~j, k = 1, \ldots, p.
\end{equation}
However, the network constructed from $\widehat{G}_{\text{GRADE}}$ tends to contain a large number of spurious edges, as illustrated in Figure~\ref{fig:noisy}. 

To address this issue, we develop a data-driven threshold selection procedure that substantially improves the accuracy of the reconstructed regulatory network. Let $\mathcal{T}_j$ denote the set of estimated strengths of all non-zero regulatory effects from every node to node $j$, including the self-regulatory effect.
\[
\mathcal{T}_j=\{\ \|\widehat{\boldsymbol{\theta}}_{jk}\|_2 : k=1,\ldots,p, \quad \|\widehat{\boldsymbol{\theta}}_{jk}\|_2 > 0 \}.
\]
Then the set
\[
\mathcal{T}=\bigcup_{j = 1}^{p} \mathcal{T}_j,
\]
contains all non-zero edge strengths in the entire network.

The estimated regulatory strengths, $\|\widehat{\boldsymbol{\theta}}_{jk}\|_2$, exhibit substantial variation across nodes, leading to considerable differences in the scales of the sets $\mathcal{T}_j$. To facilitate comparisons between nodes and preserve the strongest regulatory edges at each node, $\|\widehat{\boldsymbol{\theta}}_{jk}\|_2$ is normalized independently for each node $j=1,\ldots,p$. Define
\[
\widetilde{\mathcal{T}}_j= \left\{\dfrac{\|\widehat{\boldsymbol{\theta}}_{jk}\|_2-\min(\mathcal{T}_j)}{\max(\mathcal{T}_j)-\min(\mathcal{T}_j)}:~ k = 1,\ldots, p, \quad \|\widehat{\boldsymbol{\theta}}_{jk}\|_2 > 0 \right\},
\]
and let
\[
\widetilde{\mathcal{T}}=\bigcup_{j=1}^{p}\widetilde{\mathcal{T}}_j.
\]
Spurious edges typically have relatively small estimated regulatory strengths. However, it is unclear how small an estimated strength should be for the corresponding edge to be discarded. Directly screening the elements of $\widetilde{\mathcal{T}}$ lacks robustness because neighboring elements often differ only slightly, making the separation highly sensitive to finite-sample fluctuations, smoothing variation, estimation uncertainty, and other sources of random errors. To obtain a more reliable threshold, we first sort and cluster the elements of $\widetilde{\mathcal{T}}$ according to their estimated strengths. The clustering procedure partitions the elements of $\widetilde{\mathcal{T}}$ into groups such that elements within the same group have similar strengths, whereas those in different groups are well separated. Consequently, a threshold placed between adjacent groups corresponds to larger and more robust strength differences than the one placed between individual elements. 

Specifically, we apply the Gaussian Mixture Model (GMM) to the elements of $\widetilde{\mathcal{T}}$ to cluster the edges according to their estimated strengths~\citep{fraley_model-based_2002}. Because the number of clusters is unknown, a sequence of candidate values is explored, and the optimal value $K$ is selected using the BIC~\citep{scrucca_mclust_2016}. Let $\mathcal{C}_1, \ldots, \mathcal{C}_K$ denote the resulting clusters, which are sorted in a decreasing order so that $\mathcal{C}_1$ has the largest strength mean and $\mathcal{C}_K$ the smallest. To identify the most pronounced separation among the sorted clusters, we apply a maximum-ratio criterion. Specifically, for each pair of adjacent clusters $\mathcal{C}_{\mathscr{k}}$ and $\mathcal{C}_{\mathscr{k}+1}$, we compute the drop ratio
$\min(\mathcal{C}_{\mathscr{k}})/\max(\mathcal{C}_{\mathscr{k}+1}),$
which quantifies the relative separation level between each neighboring cluster pair. The pair with the largest drop ratio is regarded as the optimal partition, and its location is given by
\[
\mathscr{k}^{\star}
=
\arg\max_{\mathscr{k}}
\frac{\min(\mathcal{C}_{\mathscr{k}})}
{\max(\mathcal{C}_{\mathscr{k}+1})}, \qquad
\mathscr{k}=1,\ldots,K-1.
\]
Once $\mathscr{k}^\star$ is determined, we reconstruct the regulatory network according to the adjacency matrix $\widehat{G}_{\text{RECON}}$ defined as follows:
\begin{equation}\label{eq:g-RECON}
\widehat{G}_{\text{RECON}}[k,j] =
\begin{cases}
1, & 
\text{if } 
\|\widehat{\boldsymbol{\theta}}_{jk}\|_2 
\ge 
\min(\mathcal{C}_{\mathscr{k}^\star})\,
\bigl(\max(\mathcal{T}_j)-\min(\mathcal{T}_j)\bigr)
+ \min(\mathcal{T}_j), \\[6pt]
0, & \text{otherwise}.
\end{cases}
\end{equation}

\subsection{Interpreting the Reconstructed Network}\label{sec:method-interp}
Once the regulatory network is reconstructed, a comprehensive interpretation is essential for extracting scientific insights from real data, including the sign and strength of each regulatory edge, the identification of keystone nodes, and the topological organization of the network.

The sign of the regulatory edge, $k \rightarrow j$, is defined as either \emph{activation} (positive sign) or \emph{inhibition} (negative sign). Activation refers to positive relationships, including mutualism, cooperation, and synergism, where one state variable promotes the growth or function of another. In contrast, inhibition refers to negative relationships, such as predation, parasitism, and competition, where one state variable suppresses the growth or function of another. However, the regulatory effect of $k \rightarrow j$ is described by a dynamic function of $t$, i.e., $f_{jk}(X_k(t;\boldsymbol{\theta}))$, which can be estimated by
\begin{equation}\label{eq:f-hat}
\widehat{f}_{jk}(\widehat{X}_k(t;\mathbf{h},\boldsymbol{\theta})) = \boldsymbol{\psi}(\widehat{X}_k(t;\mathbf{h},\boldsymbol{\theta}))^\top \widehat{\boldsymbol{\theta}}_{jk}, \quad t\in [0,1], \quad j,k = 1,\ldots, p,
\end{equation}
where $\widehat{X}_k(t;\mathbf{h},\boldsymbol{\theta})= \frac{1}{R} \sum_{r = 1}^{R} \widehat{X}_k^{(r)}(t;\mathbf{h},\boldsymbol{\theta})$.
These functions may fluctuate above and below the zero line over the study period, complicating the determination of their signs. Therefore, we estimate its overall effect by the sign of the area that each function covers over the entire domain,
\[
S_{jk} = \int_{0}^{1} \widehat{f}_{jk}(\widehat{X}_k(t;\mathbf{h},\boldsymbol{\theta}))\, dt.
\]
A positive value of $S_{jk}$ indicates activation, while a negative value of $S_{jk}$ indicates inhibition for each edge $k\rightarrow j$. 
 
 In the following, we introduce \emph{Degree Centrality}, \emph{Modularity}, and \emph{Betweenness Centrality} to characterize the corresponding topological structure of the reconstructed network. Degree centrality of node $j$ is measured by the number of edges connected to that node including incoming and outgoing edges, which is defined as
\[
    d_j = \sum_{k = 1}^{p} \big( \widehat{G}_{\text{RECON}}[k,j] + \widehat{G}_{\text{RECON}}[j,k] \big).
\]
A higher degree indicates that a node has more regulatory interactions with other nodes. In particular, nodes with the highest degrees often function as hubs and are considered potential keystone nodes because they exert broad influence on the overall network structure and system dynamics~\citep{berry_deciphering_2014,de_vries_soil_2018,layeghifard_disentangling_2017}.

Modularity ($H$) of a network is a metric ranging from $-1$ to $1$ that quantifies the degree to which a network is partitioned into communities with relatively few connections between them. High modularity reflects increased community resilience, as disturbances such as node removal are typically confined within individual modules, thereby reducing the overall impact on the network~\citep{lima-mendez_determinants_2015,faust_microbial_2012,hernandez_environmental_2021}. Modularity of the reconstructed network is defined by,
\[
H = \frac{1}{2D}\sum_{j =1}^{p}\sum_{k = 1}^p \big( \widehat{G}_{\text{RECON}}[j,k] - \frac{d_j d_k}{2D} \big) \delta(c_j,c_k),
\]
where $D = \frac{1}{2}\sum_{j=1}^{p} d_j$. The variables $c_j$ and $c_k$ denote the modules to which nodes $j$ and $k$ belong, respectively. The function $\delta$ denotes the Kronecker delta, such that $\delta(c_j,c_k) = { \mathbbm{1}}_{\{c_j = c_k\}}$. In our analysis, the values of $c_1,\ldots,c_p$ are identified by applying the Louvain community detection method~\citep{blondel_fast_2008}. 

Betweenness centrality of a node measures the extent to which this node lies on the shortest paths connecting other pairs of nodes in the network. A shortest path between two nodes is defined as the one containing the minimum number of edges~\citep{freeman_set_1977, brandes_faster_2001}. Formally, betweenness centrality of node $j$ is defined as,
\[
\varrho(j) = \sum_{\substack{k_1,k_2 = 1 \\k_1,k_2\neq j}}^p \frac{SP_{k_1,k_2}(j)}{SP_{k_1,k_2}}.
\]
Here, $SP_{k_1,k_2}$ denotes the total number of the directed shortest paths from node $k_1$ to $k_2$ strictly following the edge directions, and $SP_{k_1,k_2}(j)$ represents the number of these paths that pass through node $j$. Nodes with high betweenness centrality serve as bridges between otherwise separate regions of the network. Such nodes are often termed as \emph{gatekeepers}, since their removal can lead to significant network fragmentation. Algorithm~\ref{alg:RECON} summarizes the complete workflow of the proposed RECON approach.

\begin{algorithm}[!ht]
\caption{Complete Workflow of the Proposed RECON Approach}\label{alg:RECON}
{\small
\begin{algorithmic}
\State \textbf{Input:} Observed data $\{Y^{(r)}_{ij}; \quad r = 1,\ldots,R, \quad i = 1,...,n_r, \quad j = 1,\ldots,p\}$ ; the discretely observed noisy observations of the $p$ state trajectories $X^{(r)}_1(t;\boldsymbol{\theta}),\ldots,X^{(r)}_p(t;\boldsymbol{\theta})$ for subject $r$ at time $t_i$ as given in Equation~\eqref{eq:obs-rep} or Equation~\eqref{eq:obs-pace}\\

\For{$r = 1$ to $R$}
    \For{$j = 1$ to $p$}
        \If{sampling is dense and regular}
            \State Obtain the initial smooth $\widetilde{X}^{(r)}_j(t;\mathbf{h})$ by local polynomial smoothing, as in Equation~\eqref{eq:locpoly-r}
        \Else{ \textbf{if} sampling is sparse and irregular}
            \State Obtain the initial smooth $\widetilde{X}^{(r)}_j(t;\mathbf{h})$ by PACE, as in~\eqref{eq:pace-r} and using AIC to determine the number of principal components
        \EndIf
     \State The bandwidths $\mathbf{h}$ are chosen by GCV
        
    \EndFor
    
    \State Estimate each regulatory function $f_{jk}(\cdot)$ non-parametrically from the additive ODE structure in Equation~\eqref{eq:additive-ode} by a B-spline basis $\boldsymbol{\psi}(\cdot) = (\psi_1(\cdot),\ldots,\psi_M(\cdot))^\top$ as in Equation~\eqref{eq:basis}

    \State Substitute $\widetilde{\mathbf{X}}^{(r)}(t;\mathbf{h}) = \bigl(\widetilde{X}^{(r)}_1(t;\mathbf{h}),\ldots,\widetilde{X}^{(r)}_p(t;\mathbf{h})\bigr)^\top$ into the integrated additive ODE structure to obtain the final estimate $\widehat{\mathbf{X}}^{(r)}(t;\mathbf{h},\boldsymbol{\theta})$, as in Equation~\eqref{eq:tildeX}
\EndFor

\For{$j = 1$ to $p$}
    \State Select the group LASSO regularization parameter $\lambda_{n,j}$ by the BIC, and obtain the estimates $\widehat{\boldsymbol{\theta}}_{j} = \{\widehat{\boldsymbol{\theta}}_{jk} :k=1,\ldots,p\}$, following Equation~\eqref{eq:opt-rep}
\EndFor

\State Normalize the estimated interaction strengths $\|\widehat{\boldsymbol{\theta}}_{jk}\|_2$ separately at each node $j$, obtaining $\widetilde{\mathcal{T}}$

\State Cluster the elements of $\widetilde{\mathcal{T}}$ using a Gaussian Mixture Model, selecting the number of clusters $K$

\State Apply the maximum-ratio criterion to the sorted clusters to identify the optimal cluster boundary $\mathscr{k}^\star$

\State Construct $\widehat{G}_{\text{RECON}}$ by retaining edges with strength above the threshold implied by $\mathscr{k}^\star$, as in Equation~\eqref{eq:g-RECON}

\For{$j = 1$ to $p$}
    \For{$k = 1$ to $p$}
        \State Estimate each regulatory function $\widehat{f}_{jk}(\widehat{X}(t;\mathbf{h},\boldsymbol{\theta}))$ from $\widehat{\boldsymbol{\theta}}_{jk}$, as in Equation~\eqref{eq:f-hat}, and obtain the edge sign $S_{jk}$ by integrating $\widehat{f}_{jk}$ over $[0,1]$
    \EndFor
\EndFor

\\
\State \textbf{Output:} Adjacency matrix $\widehat{G}_{\text{RECON}}$, Signs of the edges $S_{jk}$ for $ j,k = 1,\ldots,p$, Modularity $H$ of the reconstructed network, Degree $d_j$ and Betweenness $\varrho(j)$ of each node $j = 1,\ldots,p$.

\end{algorithmic}
}
\end{algorithm}

\section{Simulation Studies}\label{sec:sim}
In this section, we evaluate the performance of RECON through simulation studies. We consider three ODE systems corresponding to three different topological structures of networks and two simulation designs. Simulation~1-I refers to the first ODE system under Design~I, whereas Simulation~2-II refers to the second ODE system under Design~II, etc. In Simulations~1-I, 1-II, and~3-I, we set $p=8,16,30$, whereas in Simulations~2-I and~2-II, we set $p=9,18$. The number of subjects is fixed at $R=200$ throughout all simulation settings.

Let $B \subset \{1, \ldots, p\}$ be a subset of indices, and let 
$\mathbf{X}_B(t;\boldsymbol{\theta}) = \{X_b(t;\boldsymbol{\theta})\,;\, b \in B\}$ denote the trajectories of the state variables indexed by $B$. We call $B$ a block if the dynamics of every variable inside $\mathbf{X}_B$ depend only on other variables that are also within $\mathbf{X}_B$, that is
\[
f_b(\mathbf{X}(t;\boldsymbol{\theta}),\boldsymbol{\theta}) = 
f_b(\mathbf{X}_B(t;\boldsymbol{\theta}),\boldsymbol{\theta}) \quad 
\text{for all } b \in B.
\]
Additionally, no variable outside $\mathbf{X}_B$ is influenced by any variable inside $\mathbf{X}_B$:
\[
\frac{\partial f_{b'}}{\partial X_b} \equiv 0 \quad 
\text{for all } b \in B,\ b' \notin B.
\] 
If the ODE system in Equation~\eqref{eq:ode} contains a block $B$, the true network should contain a module whose node set is exactly $B$. Let $Q$ denote the number of blocks in the ODE system in Equation~\eqref{eq:ode}. In Simulations~1 and~2, we consider linear ODE systems with $Q = p/2$ and $ Q = p/3$ blocks, respectively. In Simulation~3, we consider a nonlinear ODE system with $Q = p/2$ blocks.

\subsection{Simulation Design I}\label{sec:sim-design}

We adopt the following simulation design suggested by~\citet{chen_network_2017}. Since the true parameter $\boldsymbol{\theta}^*$ is already specified by the ODE system, the corresponding $p \times p$ binary adjacency matrix $G^*$, representing the true network, is then constructed directly from $\boldsymbol{\theta}^*$. For each subject $r \in \{1, \ldots, R\}$, the initial condition $\mathbf{X}^{(r)}(0)$ is drawn from a simulation-specific distribution to induce subject-specific trajectory variation. Since the ODE system is fully specified in the simulation studies, its true trajectories $\mathbf{X}^{(r)}(t;\boldsymbol{\theta}^{*})$ are obtained by numerically solving the ODE using the Euler method:
\begin{equation}\label{eq:euler}
\mathbf{X}^{(r)}(s_{i+1};\boldsymbol{\theta}^*)
=
\mathbf{X}^{(r)}(s_i;\boldsymbol{\theta}^*)
+
\mathbf{f}\bigl(\mathbf{X}^{(r)}(s_i;\boldsymbol{\theta}^*),\boldsymbol{\theta}^*\bigr)\Delta s.
\end{equation}
To ensure accurate numerical solutions, we apply a very fine time grid with step size $\Delta s=0.001$, where $s_i=i\Delta s$ for $i=1,\ldots,1000$. After obtaining the smooth trajectory $\mathbf{X}^{(r)}(t;\boldsymbol{\theta}^{*})$, we evaluate it at a common and dense set of time points $t_i$ to mimic real data where observations are measured discretely. Specifically, we choose $n=5p$ equally spaced time points, and obtain $\mathbf{X}^{(r)}(t_i;\boldsymbol{\theta}^{*})$ at $t_i=i/n,~ i=1,\ldots,n$. The discrete observations $\mathbf{Y}^{(r)}_i$ are then generated by adding Gaussian noise according to Equation~\eqref{eq:obs-rep}, where each component $\varepsilon^{(r)}_{ij}$ is independently generated from $\mathscr{N}(0,1)$.

Now we complete the data-generating process for Simulation Design~I. In the subsequent analysis, only the noisy observations $\{\mathbf{Y}^{(r)}_i\}$ are used, while both the true trajectories $\mathbf{X}^{(r)}(t;\boldsymbol{\theta}^{*})$ and the true parameter $\boldsymbol{\theta}^{*}$ are treated as unknown, as in real data applications. The RECON workflow described in Algorithm~\ref{alg:RECON} is then applied to $\{\mathbf{Y}^{(r)}_i\}$, yielding the reconstructed adjacency matrix $\widehat{G}_{\text{RECON}}$, which is compared with the true network $G^*$ and with $\widehat{G}_{\text{GRADE}}$ obtained from Equation~\eqref{eq:g-bic}.

\subsection{Simulation Design II}\label{sec:sim-ic} Consider a setting with $N_m$ repeated measurements of scalar predictor vectors, each having $p$ components, and $N_m$ repeatedly measured functional response trajectories for each subject $r=1,\ldots,R$. Let $V_1,\ldots,V_p$ denote the $p$ scalar predictor variables. For subject $r$, let $\mathbf V_m^{(r)}=\bigl(V_{m1}^{(r)},\ldots,V_{mp}^{(r)}\bigr)\in\mathbb R^p,~m=1,\ldots,N_m,$ denote the predictor vector at the $m$th repeated measurement, whose $j$th component $V_{mj}^{(r)}$ is the observed value of predictor variable $V_j$. Let $\mathbf W_m^{(r)}=\bigl(W_{m1}^{(r)},\ldots,W_{mn}^{(r)}\bigr) \in\mathbb R^n$ denote the $m$th repeatedly and discretely measured functional response evaluated at the common time points
$t_1,\ldots,t_n$
shared across all subjects. The observation model is
\begin{equation}\label{eq:func-obs}
W_{mi}^{(r)}
=
U_m^{(r)}(t_i)
+
\xi_{mi}^{(r)},
\qquad
i=1,\ldots,n,\;
m=1,\ldots,N_m,\;
r=1,\ldots,R,
\end{equation}
where
$U_m^{(r)}(t)$
is the underlying smooth functional trajectory for the $m$th replicate of subject $r$. The observation errors
$\xi_{mi}^{(r)}$
are assumed to be independently and identically distributed across subjects, repeated measurements, and time points. Depending on the sampling scenario, the discretely measured $\mathbf{W}_m^{(r)}$ should first be smoothed using the procedures described in Equations~\eqref{eq:locpoly-r} or~\eqref{eq:pace-r} to obtain estimates $\widehat{U}_m^{(r)}(t;\mathbf{h})$ of the underlying smooth trajectories $U_m^{(r)}(t)$. 

The latent function-on-scalar regression model is then formulated for the unobserved trajectories as

\begin{equation}\label{eq:funreg}
U_m^{(r)}(t)
=
\mathbf V_m^{(r)}
\boldsymbol\beta^{(r)}(t;\boldsymbol\theta)
+
\mathscr e_m^{(r)}(t),
\qquad
t\in[0,1], \quad r = 1, \ldots, R,
\end{equation}
where
\[
\boldsymbol\beta^{(r)}(t;\boldsymbol\theta)
=
\bigl(
\beta_1^{(r)}(t;\boldsymbol\theta),
\ldots,
\beta_p^{(r)}(t;\boldsymbol\theta)
\bigr)^{\mathsf T}
\]
denotes the subject-specific vector of latent functional coefficient trajectories, which is shared across all the $N_m$ repeated measurements within subject $r$. The error process
$\mathscr e_m^{(r)}(t)$
has mean zero and is assumed to be independent across subjects and repeated measurements, and independent of the predictor vectors
$\mathbf V_m^{(r)}$.

Combining Equations~\eqref{eq:func-obs}
and~\eqref{eq:funreg},
the noisy observations satisfy
\begin{equation}\label{eq:fun-reg-obs}
W_{mi}^{(r)}
=
\sum_{j=1}^p
V_{mj}^{(r)}
\beta_j^{(r)}(t_i;\boldsymbol\theta)
+
\mathscr e_m^{(r)}(t_i)
+
\xi_{mi}^{(r)},
\qquad
i=1,\ldots,n,\;
m=1,\ldots,N_m,\;
r=1,\ldots,R.
\end{equation}

For each subject $r$, define the $N_m\times p$ predictor matrix
\[
\mathbf{V}^{(r)} =
\begin{pmatrix}
\mathbf{V}_1^{(r)}\\
\vdots\\
\mathbf{V}_{N_m}^{(r)}
\end{pmatrix}.
\]
We set that $N_m\ge p$ and that $\mathbf{V}^{(r)}$ has full column rank. Together with the mean-zero error assumptions, this condition ensures that the subject-specific coefficient trajectories $\boldsymbol\beta^{(r)}(t;\boldsymbol\theta)$ are identifiable.


To characterize the regulatory effects of the scalar predictors on the functional response, we take the $p$ scalar predictors $V_1,\ldots,V_p$ as the nodes of the network. The network edges are determined by treating the corresponding subject-specific latent functional coefficient trajectories,
$\boldsymbol{\beta}^{(r)}(t;\boldsymbol{\theta}^*)$, as the state variables whose dynamics are governed by the ODE system
\begin{equation}\label{eq:ode-beta}
\boldsymbol{\beta}'(t;\boldsymbol{\theta}^*)
=
\mathbf{f}\bigl(\boldsymbol{\beta}(t;\boldsymbol{\theta}^*),\boldsymbol{\theta}^*\bigr),
\qquad
t\in[0,1].
\end{equation}
We assume that all subjects share the same underlying ODE system and true parameter
$\boldsymbol{\theta}^{*}$, but differ in their initial conditions, resulting in subject-specific coefficient trajectories $\boldsymbol{\beta}^{(r)}(t;\boldsymbol{\theta}^{*})$. 

We now describe the data-generating process. For each subject $r=1,\ldots,R$ and replicate $m=1,\ldots,N_m$, the scalar predictor component, $V_{mj}^{(r)}$, is generated by independently sampling from $\mathrm{Unif}(0,1)$. Once the true ODE system is specified, a subject-specific initial condition $\boldsymbol{\beta}^{(r)}(0)$ is generated from a simulation-specific distribution, and the corresponding coefficient trajectories $\boldsymbol{\beta}^{(r)}(t;\boldsymbol{\theta}^{*})$ are obtained numerically using the Euler method described in Equation~\eqref{eq:euler}. We then set $n=5p$ and evaluate $\boldsymbol{\beta}^{(r)}(t;\boldsymbol{\theta}^{*})$ at $n$ equally spaced time points $t_1,\ldots,t_n$ for each subject $r=1,\ldots,R$. The functional regression errors $\mathscr e_m^{(r)}(t_i)$ and observation errors $\xi_{mi}^{(r)}$ are independently generated from $\mathscr{N}(0,\sigma^2)$ across subjects, replicates, and time points, where $\sigma=0.5/\sqrt{2}$. Finally, the discretely observed functional responses $W_{mi}^{(r)}$ are generated according to Equation~\eqref{eq:fun-reg-obs}.

Once the data-generating process is complete, the subsequent analysis proceeds using only the observed functional response and scalar predictor replications,\( \{(\mathbf W_m^{(r)},\mathbf V_m^{(r)});  m=1,\ldots,N_m \} \), yielding the estimated subject-specific latent coefficient trajectories $\widetilde{\boldsymbol{\beta}}^{(r)}(t)$, which serve as the initial smooth estimates. Unlike Simulation Design~I, no additional smoothing step via Equations~\eqref{eq:locpoly-r} or~\eqref{eq:pace-r} is required because the latent functional regression model \eqref{eq:funreg} directly estimates the smooth coefficient trajectories. The remaining analysis follows Algorithm~\ref{alg:RECON}.

In Simulation Design~I, the noisy measurements $\mathbf Y_i^{(r)}$ provide discrete observations of the $p$ ODE state trajectories $\mathbf X^{(r)}(t;\boldsymbol\theta)$, from which the network is reconstructed directly. Consequently, the reconstructed network characterizes the regulatory relationships among the $p$ observed state variables, which also serve as the network nodes. Simulation Design~II is considerably more challenging because the ODE state variables are no longer directly observed. Instead, they appear as latent coefficient trajectories $\boldsymbol{\beta}^{(r)}(t;\boldsymbol\theta)$ embedded within a latent function-on-scalar regression model, introducing an additional source of estimation error. Furthermore, because the network is reconstructed from the latent coefficient trajectories
$\boldsymbol{\beta}^{(r)}(t;\boldsymbol\theta)$ rather than directly from the scalar predictors $\mathbf V_k^{(r)}$, the reconstructed network characterizes the dynamic regulatory effects of the scalar predictors on the functional response rather than the relationships among the scalar predictors themselves.


\subsection{Assessment Criteria}\label{sec:sim-eval}

Let $\widehat{G}$ (i.e., $\widehat{G}_{\text{GRADE}}$ by Equation~\eqref{eq:g-bic} or $\widehat{G}_{\text{RECON}}$ by Equation~\eqref{eq:g-RECON}) denote an estimated adjacency matrix. Each entry of $\widehat{G}[k,j]$ and of the true adjacency matrix $G^{*}[k,j]$ is either $0$ or $1$, indicating the absence or presence of a directed edge $k \to j$, where $k,j \in \{1,\ldots,p\}$. For each ordered pair $(k,j)$, if both $\widehat{G}[k,j] = 1$ and $G^*[k,j] = 1$, the edge $k \to j$ is correctly recovered; if both equal $0$, the absence of the edge is correctly identified; and if they disagree, the edge is either a spurious one or a missed true edge. The accuracy of the reconstructed network can therefore be evaluated using performance measures adopted from the binary classification literature. We report three criteria: true positives (TP), i.e. the number of correctly recovered edges; false positives (FP), i.e. the number of spurious edges; and false negatives (FN), i.e. the number of missed edges, defined as,
\begin{align*}
\mathrm{TP}(\widehat{G}) &= \mathbf{1}^\top \bigl(\widehat{G} \odot G^{*}\bigr)\,\mathbf{1}, \\
\mathrm{FP}(\widehat{G}) &= \mathbf{1}^\top \bigl(\widehat{G} \odot (\mathbf{1}\mathbf{1}^\top - G^{*})\bigr)\,\mathbf{1}, \\
\mathrm{FN}(\widehat{G}) &= \mathbf{1}^\top \bigl(G^{*} \odot (\mathbf{1}\mathbf{1}^\top - \widehat{G})\bigr)\,\mathbf{1},
\end{align*}
where $\odot$ denotes the elementwise (Hadamard) product and $\mathbf{1} = (1,\ldots,1)^\top\in \mathbb{R}^{p}$. 
From TP and FP, we compute the true positive rate (TPR), false positive rate (FPR), and the area under the ROC curve (AUC-ROC) as \citep{hanley_meaning_1982,matthews_comparison_1975},
\begin{align*}
\mathrm{TPR}(\widehat{G}) &= \frac{\mathbf{1}^\top(\widehat{G} \odot G^{*})\,\mathbf{1}}{\mathbf{1}^\top G^{*}\,\mathbf{1}}, \\
\mathrm{FPR}(\widehat{G}) &= \frac{\mathbf{1}^\top(\widehat{G} \odot (\mathbf{1}\mathbf{1}^\top - G^{*}))\,\mathbf{1}}{\mathbf{1}^\top(\mathbf{1}\mathbf{1}^\top - G^{*})\,\mathbf{1}}, \\
\text{AUC-ROC} &= \frac{1+\mathrm{TPR}(\widehat{G})-\mathrm{FPR}(\widehat{G})}{2}.
\end{align*}
Here, smaller TP or larger FN indicates possible model misclassification, larger FP indicates greater number of spurious edges generated in the network~\citep{stehman_selecting_1997}. An AUC-ROC of $0.5$ represents random guessing; higher values ($\geq 0.7$ fair, $\geq 0.8$ good, $\geq 0.9$ strong) indicate better network reconstruction accuracy.

\subsection{Simulation 1: Linear system with two-way blocks}\label{sec:sim1}

This simulation evaluates network reconstruction using an ODE structure consisting of $Q = p/2$ number of two-way blocks with no between-block interactions. The corresponding ODE system is given below.

For $q = 1, \ldots, Q$, 
\begin{equation}\label{eq:sim1-ode}
\begin{cases}
X'_{2q-1}(t) = 2q\pi\, X_{2q}(t), \\[4pt]
X'_{2q}(t)   = -2q\pi\, X_{2q-1}(t),
\end{cases}
\qquad t \in [0,1],
\end{equation}
which generates modules, each containing interacting variable pairs $(X_{2q-1}, X_{2q})$ and no connection with other modules.

\subsubsection*{Simulation 1-I}\label{sec:sim1a}
In this subsection, we apply Design~I to the ODE system defined in Equation~\eqref{eq:sim1-ode}. For each subject $r = 1, \ldots, R$, the initial conditions are randomly generated as
\begin{equation}
\left\{
\begin{aligned}
X_{2q-1}^{(r)}(0) &= \sin(\zeta_q^{(r)}), \\[4pt]
X_{2q}^{(r)}(0)   &= \cos(\zeta_q^{(r)}),
\end{aligned}
\right.
\end{equation}
where $\zeta_q^{(r)} \sim \mathscr{N}(0,1)$. The true network resulting from the ODE system in Equation~\eqref{eq:sim1-ode} has adjacency matrix
\begin{equation}\label{eq:sim1-G}
    G^*[k,j] =
        \begin{cases}
        1, & \text{if } (k,j) = (2q,2q-1), (2q-1,2q),\\
        0, & \text{otherwise,}
        \end{cases}
\end{equation}
for $j,k = 1,\ldots,p$ and $q = 1,\ldots,Q$.

\begin{figure}[t]
\centering

\begin{subfigure}{1.75in}

\includegraphics[width=\linewidth]{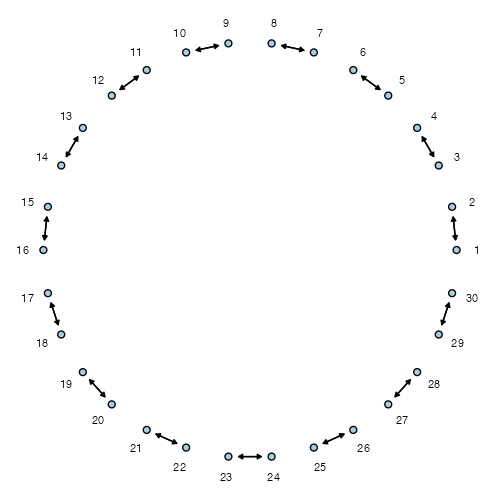}
\caption{TRUTH}
\label{fig:sim1-truth}
\end{subfigure}
\hfill
\begin{subfigure}{1.75in}

\includegraphics[width=\linewidth]{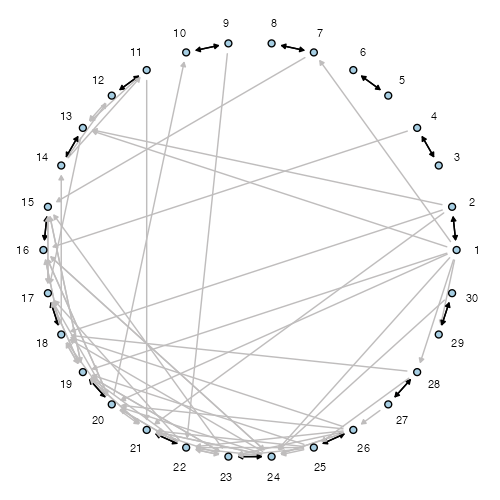}
\caption{GRADE}
\label{fig:sim1-bic}
\end{subfigure}
\hfill
\begin{subfigure}{1.75in}
\includegraphics[width=\linewidth]{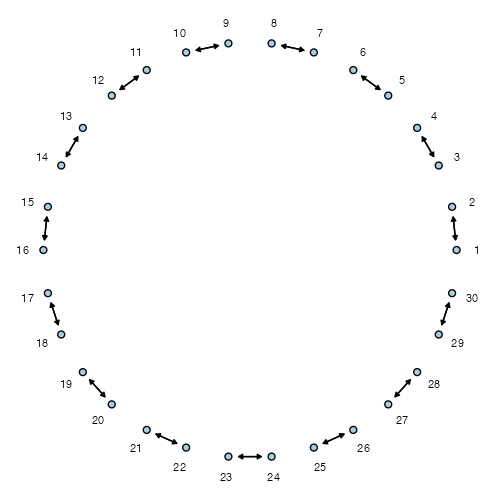}
\caption{RECON}
\label{fig:sim1-nc}
\end{subfigure}
\hfill
\caption{Comparison of the true and reconstructed networks for Simulation~1-I with $p=30$. Nodes are represented by dots, true edges by black arrows (\textcolor[HTML]{000000}{$\rightarrow$}), and spurious edges by gray arrows (\textcolor[HTML]{BDBDBD}{$\rightarrow$}). (a) True network. (b) The network constructed by GRADE, containing 70 spurious edges. (c) The network reconstructed by RECON, achieving $100\%$ accuracy.}
\label{fig:sim1}
\end{figure}

Figure~\ref{fig:sim1-truth} presents the true network, in which edges connect only adjacent nodes on the circle, reflecting the underlying two-way block structure. Figure~\ref{fig:sim1-bic} shows the network reconstructed using GRADE, which contains $70$ spurious edges connecting non-adjacent nodes across the circle. In contrast, Figure~\ref{fig:sim1-nc} shows that RECON recovers all $30$ true edges without introducing any spurious edges. Table~\ref{tab:sim1} further confirms that RECON achieves an AUC-ROC of $1.00$ across all three network sizes, indicating perfect network reconstruction. For example, when $p=30$, GRADE produces $70$ false positives, whereas RECON produces none. RECON reconstructs the true network with $100\%$ accuracy for all three values of $p$ in Simulation~1-I.

\begin{table}[t]
\caption{Comparison of network reconstruction accuracy by GRADE and RECON for Simulation~1-I across network sizes $p=8,16,30$. TP: true positives; FP: false positives; FN: false negatives.}
\label{tab:sim1}
\centering
\begin{tabular}{|c|c|c|c|c|c|c|}
\hline
$p$ & Method & AUC-ROC & TP & FP & FN  \\
\hline
8  & GRADE   & 0.95 & 8  & 5  & 0  \\
   & RECON    & \textbf{1.00} & 8  & 0  & 0 \\
\hline
16 & GRADE   & 0.98 & 16 & 7  & 0  \\
   & RECON    & \textbf{1.00} & 16 & 0  & 0  \\
\hline
30 & GRADE   & 0.96 & 30 & 70 & 0 \\
   & RECON    & \textbf{1.00} & 30 & 0  & 0 \\
\hline
\end{tabular}
\end{table}

\subsubsection*{Simulation 1-II}\label{sec:sim1-ib}
In this subsection, we still adopt the ODE system defined in Equation~\eqref{eq:sim1-ode} but following Design~II. Therefore, the true network has the same adjacency matrix $G^*$ as defined in Equation~\eqref{eq:sim1-G}. However, the ODE structure is set up for the latent functional regression coefficients $\boldsymbol{\beta}(t;\boldsymbol{\theta})$ in Equation~\eqref{eq:funreg} rather than for the state variables $\mathbf{X}(t;\boldsymbol{\theta})$ directly, and the network is reconstructed using $\boldsymbol{\beta}(t;\boldsymbol{\theta})$. 

\begin{figure}[tb]
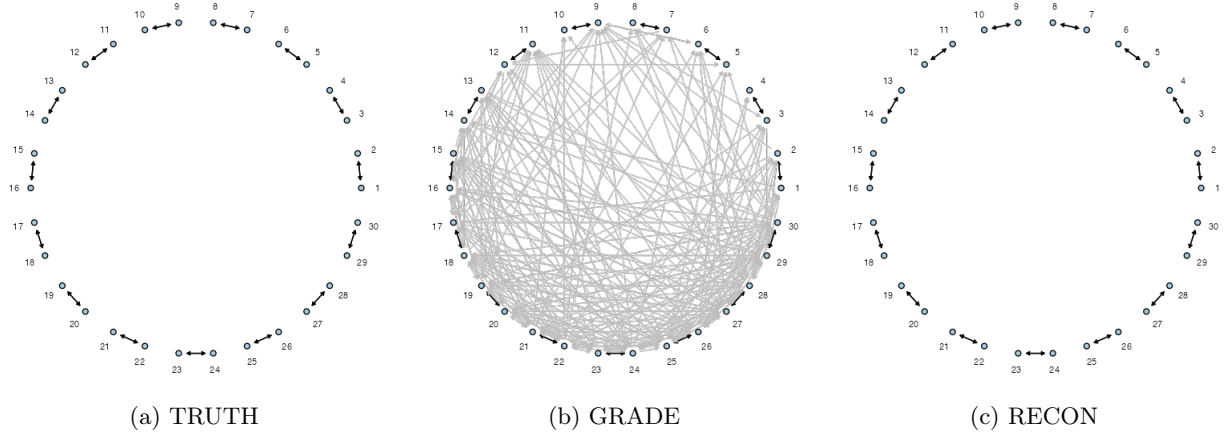

\centering

\begin{subfigure}{0.32\linewidth}

\includegraphics[width=\linewidth]{figures/sim4/true.png}
\caption{TRUTH}
\label{fig:sim1-ic-truth}
\end{subfigure}
\hfill
\begin{subfigure}{0.32\linewidth}

\includegraphics[width=\linewidth]{figures/sim4/bic.png}
\caption{GRADE}
\label{fig:sim1-ic-bic}
\end{subfigure}
\hfill
\begin{subfigure}{0.32\linewidth}

\includegraphics[width=\linewidth]{figures/sim4/nc.png}
\caption{RECON}
\label{fig:sim1-ic-nc}
\end{subfigure}
\hfill
\caption{Comparison of the true and reconstructed networks for Simulation~1-II with $p=30$. Nodes are represented by dots, true edges by black arrows (\textcolor[HTML]{000000}{$\rightarrow$}), and spurious edges by gray arrows (\textcolor[HTML]{BDBDBD}{$\rightarrow$}). (a) True network. (b) The network constructed by GRADE, containing 239 spurious edges. (c) The network reconstructed by RECON, achieving 100\% accuracy.}
\label{fig:sim1-ic}
\end{figure}

Figure~\ref{fig:sim1-ic-truth} displays the true network of Simulation~1-II, which consists of $15$ modules of two nodes each, with two within-module edges and no between-module edges. Figure~\ref{fig:sim1-ic-bic} shows the network reconstructed by GRADE, which produces $239$ spurious between-module edges. Figure~\ref{fig:sim1-ic-nc} shows the network reconstructed by RECON, which recovers all $30$ true edges and produces $0$ spurious edges. Table~\ref{tab:sim1-ic} shows that RECON achieves an AUC-ROC of $1.0000$ across all three network sizes, while GRADE's AUC-ROC drops to $0.8577$ at $p = 30$. At $p = 30$, GRADE produces $239$ false positives while RECON produces $0$. RECON reconstructs the true network achieving $100\%$ accuracy across all three network sizes in Simulation~1-II.

\begin{table}[b]
\caption{Comparison of network reconstruction accuracy by GRADE and RECON for Simulation~1-II across network sizes $p=8,16,30$. TP: true positives; FP: false positives; FN: false negatives.}
\label{tab:sim1-ic}
\centering
\begin{tabular}{|c|c|c|c|c|c|c|}
\hline
$p$ & Method & AUC-ROC & TP & FP & FN \\
\hline
8 &
  GRADE  & 0.9896 & 8  & 1   & 0 \\

 & RECON    & \textbf{1.0000} & 8  & 0   & 0 \\
\hline
16 &
  GRADE   & 0.9821 & 16 & 8   & 0 \\
 & RECON    & \textbf{1.0000} & 16 & 0   & 0 \\
\hline
30 &
  GRADE   & 0.8577 & 30 & 239 & 0 \\
 & RECON    & \textbf{1.0000} & 30 & 0   & 0 \\
\hline
\end{tabular}
\end{table}

\subsection{Simulation 2: Linear system with three-way blocks}\label{sec:sim2}

This simulation assesses the network reconstruction accuracy when the underlying ODE structure consists of $Q = p/3$ number of three-way blocks, yielding a different network structure. The corresponding ODE system is
given below.

For $q=1,\ldots,Q$,
\begin{equation}\label{eq:sim2-ode}
\left\{
\begin{aligned}
    X'_{3q-2}(t) &= -a_q X_{3q-2}(t) + c_q X_{3q}(t) + d_q t, \\
    X'_{3q-1}(t) &=  a_q X_{3q-2}(t) - w_q X_{3q-1}(t), \\
    X'_{3q}(t)   &=  w_q X_{3q-1}(t) - c_q X_{3q}(t),
\end{aligned}
\right.
\end{equation}
where $a_q,w_q,c_q\sim \text{Unif}(0,1)$ and $d_q\sim \mathrm{Unif}(1,3)$. 

\subsubsection*{Simulation 2-I}\label{subsec:sim2-1}

In this subsection, the ODE system defined in Equation~\eqref{eq:sim2-ode} is considered under Design~I. For each subject $r=1,\ldots,R$ and for each $j\in \{1,\cdots,p\}$, we randomly generate the initial conditions $X^{(r)}_{j}(0)\sim\mathrm{Unif}(1,2)$. The adjacency matrix $G^{*}$ corresponding to the true network resulting from the ODE system in Equation~\eqref{eq:sim2-ode} is defined as follows. 
For each $j,k = 1,\ldots,p $ and $q = 1,\ldots, Q$,
\begin{equation}\label{eq:sim2-G}
   G^{*}[k,j] =
    \begin{cases}
    1, & \text{if } (k,j) = (3q, 3q-2), (3q-2, 3q-1), (3q-1, 3q), \\
    1, & \text{if } k = j, \\
    0, & \text{otherwise.}
    \end{cases}
\end{equation}

\begin{figure}[t]
\centering

\begin{subfigure}{0.32\linewidth}

\includegraphics[width=\linewidth]{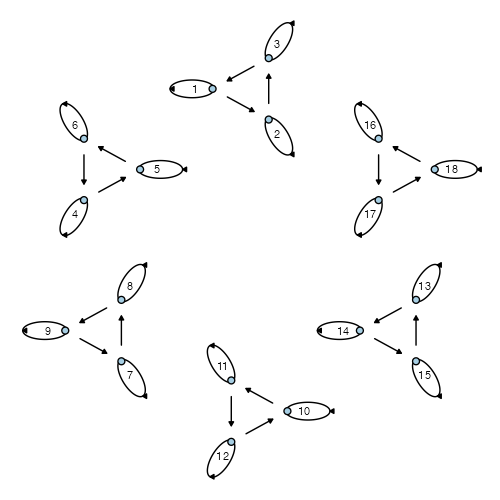}
\caption{TRUTH}
\label{fig:sim2-truth}
\end{subfigure}
\hfill
\begin{subfigure}{0.32\linewidth}

\includegraphics[width=\linewidth]{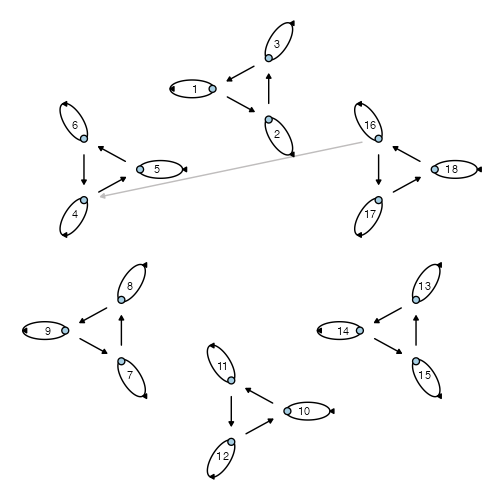}
\caption{GRADE}
\label{fig:sim2-bic}
\end{subfigure}
\hfill
\begin{subfigure}{0.32\linewidth}

\includegraphics[width=\linewidth]{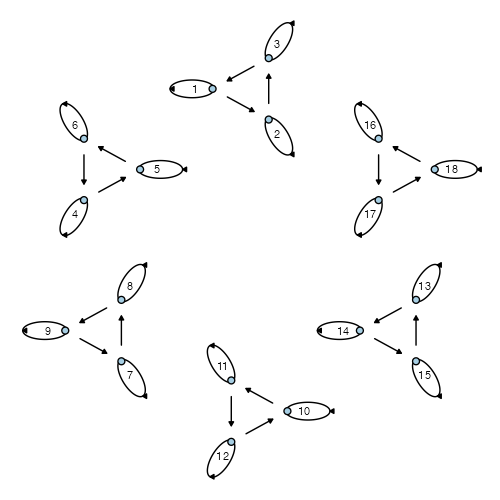}
\caption{RECON}
\label{fig:sim2-nc}
\end{subfigure}
\hfill
\caption{Comparison of the true and reconstructed networks for Simulation~2-I with $p=18$. Nodes are represented by dots, true edges by black arrows (\textcolor[HTML]{000000}{$\rightarrow$}), and spurious edges by gray arrows (\textcolor[HTML]{BDBDBD}{$\rightarrow$}). (a) True network. (b) The network constructed by GRADE, containing 1 spurious edge. (c) The network reconstructed by RECON, achieving 100\% accuracy.}
\label{fig:sim2}
\end{figure}

\begin{table}[b]
\caption{Comparison of network reconstruction accuracy by GRADE and RECON for Simulation~2-I across network sizes $p=9,18$. TP: true positives; FP: false positives; FN: false negatives.}
\label{tab:sim2}
\centering
\begin{tabular}{|c|c|c|c|c|c|}
\hline
$p$ & Method & AUC-ROC & TP & FP & FN \\
\hline
9 &
  GRADE & 0.9921 & 18 & 1 & 0 \\
& RECON & \textbf{1.0000} & 18 & 0 & 0 \\
\hline
18 &
  GRADE & 0.9983 & 36 & 1 & 0 \\
& RECON & \textbf{1.0000} & 36 & 0 & 0 \\
\hline
\end{tabular}
\end{table}

Figure~\ref{fig:sim2-truth} shows the true network of Simulation~2-I, which consists of six modules of three nodes each. Within each module, the three nodes are connected by edges forming a triangle, and each node has a self-loop. There are no edges between modules. Figure~\ref{fig:sim2-bic} shows the network reconstructed by GRADE, which produces $1$ spurious between-module edge. Figure~\ref{fig:sim2-nc} shows the network reconstructed by RECON, which recovers all $36$ true edges without producing any spurious edges. Table~\ref{tab:sim2} shows that RECON achieves an AUC-ROC of $1.0000$ at $p = 9$ and also at $p = 18$, while GRADE achieves $0.9921$ at $p = 9$ and $0.9983$ at $p = 18$. GRADE produces $1$ false positive at each network size; RECON produces $0$ false positives. In summary, the network reconstructed by RECON achieves 100\% accuracy under Simulation 2-I.

\subsubsection*{Simulation 2-II}\label{sec:sim2-2}

In this subsection, we continue to adopt the ODE system 
defined in Equation~\eqref{eq:sim2-ode} but following the Design~II. The true network has the same adjacency matrix $G^*$ as defined in Equation~\eqref{eq:sim2-G}, but assuming the ODE structure for $\boldsymbol{\beta}(t;\boldsymbol{\theta})$ instead of $\mathbf{X}(t;\boldsymbol{\theta})$.

\begin{figure}[t]
\centering

\begin{subfigure}{0.32\linewidth}

\includegraphics[width=\linewidth]{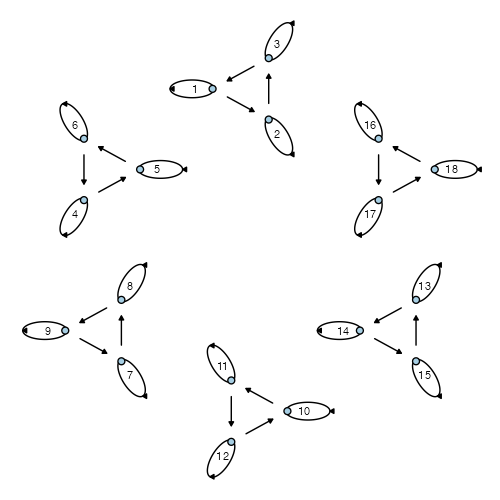}
\caption{TRUTH}
\label{fig:sim2-ic-truth}
\end{subfigure}
\hfill
\begin{subfigure}{0.32\linewidth}

\includegraphics[width=\linewidth]{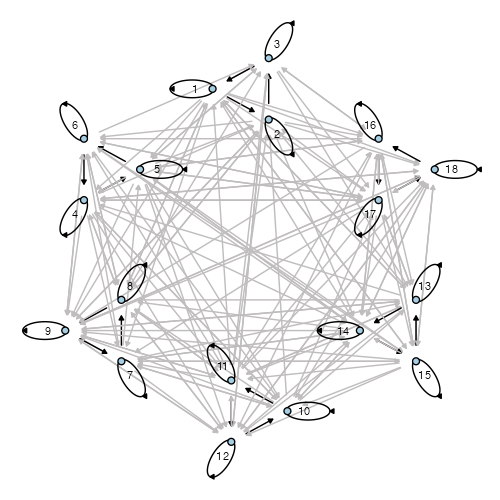}
\caption{GRADE}
\label{fig:sim2-ic-bic}
\end{subfigure}
\hfill
\begin{subfigure}{0.32\linewidth}

\includegraphics[width=\linewidth]{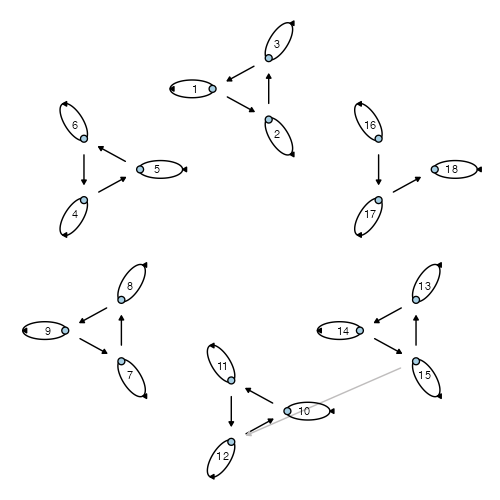}
\caption{RECON}
\label{fig:sim2-ic-nc}
\end{subfigure}
\hfill
\caption{Comparison of the true and reconstructed networks for Simulation~2-II with $p=18$. Nodes are represented by dots, true edges by black arrows (\textcolor[HTML]{000000}{$\rightarrow$}), and spurious edges by gray arrows (\textcolor[HTML]{BDBDBD}{$\rightarrow$}). (a) True network. (b) The network constructed by GRADE, containing 148 spurious edges. (c) The network reconstructed by RECON, generating only $1$ spurious edge.}
\label{fig:sim2-ic}
\end{figure}

Figure~\ref{fig:sim2-ic-truth} displays the true network for Simulation~2-II. Figure~\ref{fig:sim2-ic-bic} shows the network reconstructed by GRADE, which recovers all $36$ true edges but produces $148$ spurious edges, both within and between modules. Figure~\ref{fig:sim2-ic-nc} shows the network reconstructed by RECON, which recovers $35$ of the $36$ true edges, missing only one within-module edge $V_{18} \to V_{16}$, and produces only one spurious edge. Table~\ref{tab:sim2-ic} shows that RECON sharply reduces the FP count produced by GRADE, from $28$ to $0$ at $p = 9$ and from $148$ to $1$ at $p = 18$. At $p = 18$, RECON produces an AUC-ROC of $0.9983$, substantially improving the AUC-ROC of $0.7431$ produced by GRADE.

\begin{table}[b]
\caption{Comparison of network reconstruction accuracy by GRADE and RECON for Simulation~2-II across network sizes $p=9,18$. TP: true positives; FP: false positives; FN: false negatives.}
\label{tab:sim2-ic}
\centering
\begin{tabular}{|c|c|c|c|c|c|c|}
\hline
$p$ & Method & AUC-ROC & TP & FP & FN \\
\hline
9 &
  GRADE   & 0.7778 & 18  & 28  & 0 \\
  & RECON    & \textbf{1.0000} & 18  & 0   & 0 \\
\hline
18  &
  GRADE   & 0.7431 & 36 & 148 & 0 \\
 & RECON    & \textbf{0.9983} & 35 & 1   & 1 \\
\hline
\end{tabular}
\end{table}

\subsection{Simulation 3: Nonlinear system with two-way blocks}\label{sec:sim3}

In this simulation, we test the robustness of network reconstruction accuracy when the underlying ODE system is nonlinear. We use the Brusselator system with $Q = p/2$ blocks where each block contains $2$ nodes~\citep{prigogine_symmetry-breaking_1967}. The corresponding ODE system is
given below.

For $q = 1, \ldots, Q$ and $t \in [0, 1]$,
\begin{equation}\label{eq:sim3-ode}
\left\{
\begin{aligned}
X'_{2q-1}(t) &= a_q - (\eta_q + 1) X_{2q-1}(t) + X_{2q-1}^2(t)\, X_{2q}(t), \\
X'_{2q}(t) &= \eta_q X_{2q-1}(t) - X_{2q-1}^2(t)\, X_{2q}(t),
\end{aligned}
\right.
\end{equation}
where $a_q \sim \mathrm{Unif}(1, 2)$ and $\eta_q \sim \mathrm{Unif}(3, 5)$. The system is nonlinear due to the term $X_{2q-1}^2(t)\, X_{2q}(t)$.

\subsubsection*{Simulation 3-I}\label{subsec:sim3-1}

In this subsection, we consider the ODE system defined in Equation~\eqref{eq:sim3-ode} under the Design~I. The initial conditions $X_j^{(r)}(0)$ are drawn independently from $\mathrm{Unif}(1, 2)$ for each $j = 1, \ldots, p$ and $r = 1, \ldots, R$. The true network corresponding to this ODE system can be constructed from the adjacency matrix $G^*$ defined below. 
\begin{equation}\label{eq:sim3-G}
G^*[k, j] = \begin{cases}
1, & \text{if } (k, j) = (2q-1, 2q), (2q, 2q-1),\\
1, & \text{if } k = j, \\
0, & \text{otherwise,}
\end{cases}
\end{equation}
for each $k,j = 1,\ldots,p$ and $q = 1, \ldots, Q$.

\begin{figure}[t]
\centering

\begin{subfigure}{0.32\linewidth}

\includegraphics[width=\linewidth]{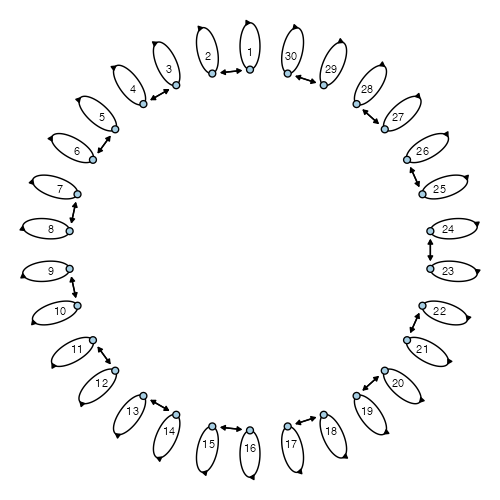}
\caption{TRUTH}
\label{fig:sim3-truth}
\end{subfigure}
\hfill
\begin{subfigure}{0.32\linewidth}

\includegraphics[width=\linewidth]{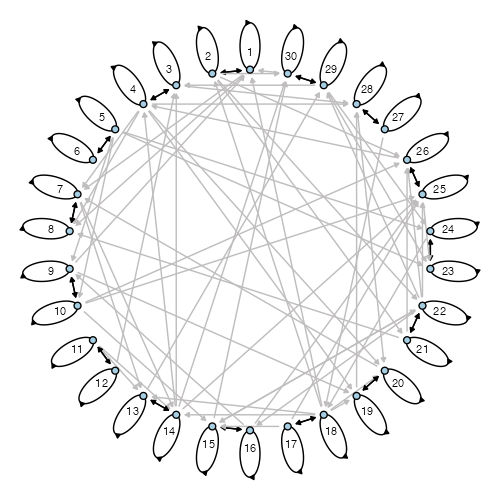}
\caption{GRADE}
\label{fig:sim3-bic}
\end{subfigure}
\hfill
\begin{subfigure}{0.32\linewidth}

\includegraphics[width=\linewidth]{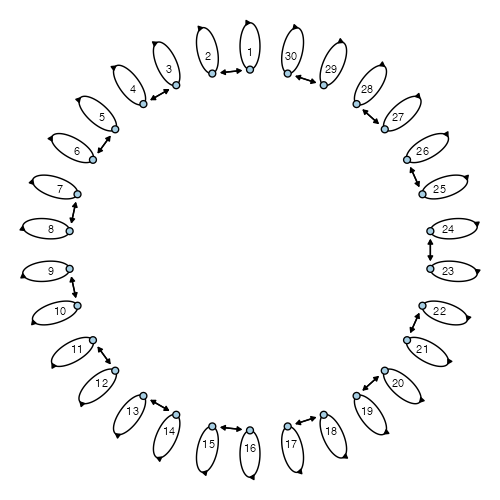}
\caption{RECON}
\label{fig:sim3-nc}
\end{subfigure}
\hfill
\caption{Comparison of the true and reconstructed networks for Simulation~3-I with $p=30$. Nodes are represented by dots, true edges by black arrows (\textcolor[HTML]{000000}{$\rightarrow$}), and spurious edges by gray arrows (\textcolor[HTML]{BDBDBD}{$\rightarrow$}). (a) True network. (b) The network constructed by GRADE, containing 61 spurious edges. (c) The network reconstructed by RECON, achieving 100\% accuracy.}
\label{fig:sim3}
\end{figure}

\begin{table}[b]
\caption{Comparison of network reconstruction accuracy by GRADE and RECON for Simulation~3-I across network sizes $p=8,16,30$. TP: true positives; FP: false positives; FN: false negatives.}
\label{tab:sim3}
\centering
\begin{tabular}{|c|c|c|c|c|c|}
\hline
$p$ & Method & AUC-ROC & TP & FP & FN \\
\hline
8 & GRADE  & 0.9375 & 16 & 6  & 0 \\
  & RECON & \textbf{1.0000} & 16 & 0  & 0 \\
\hline
16 & GRADE  & 0.9554 & 32 & 20 & 0 \\
   & RECON & \textbf{1.0000} & 32 & 0  & 0 \\
\hline
30 & GRADE  & 0.9637 & 60 & 61 & 0 \\
   & RECON & \textbf{1.0000} & 60 & 0  & 0 \\
\hline
\end{tabular}
\end{table}

Figure~\ref{fig:sim3-truth} shows the true network of Simulation~3-I, which consists of $15$ modules of two nodes each. Each module contains two edges between its nodes and a self-loop at each node, with no edges between modules. Figure~\ref{fig:sim3-bic} shows the network constructed by GRADE, which produces $61$ spurious edges. Figure~\ref{fig:sim3-nc} shows the network reconstructed by RECON, which recovers all $60$ true edges and produces $0$ spurious edges, demonstrating $100\%$ reconstruction accuracy. Table~\ref{tab:sim3} shows that RECON achieves an AUC-ROC of $1.0000$ at $p = 8, 16, 30$, while GRADE stays below $0.97$ at every network size. GRADE's false positives increase with $p$, reaching $61$ at $p = 30$, while RECON produces $0$ false positives at every size.

\section{Allo-HCT Patients Data Analysis}\label{sec:data-analysis}

\subsection{Microbiota Measurements and Preprocessing}\label{sec:prep}

We apply RECON to the MSK hospitalome dataset introduced in Section~\ref{sec:data}, which comprises longitudinal gut microbiota data from allo-HCT patients obtained using 16S rRNA sequencing. For each stool sample, the raw data consist of absolute abundance counts for all the bacterial families. Because absolute abundance counts vary substantially across samples, they are not directly comparable and can bias downstream statistical estimation. Therefore, the absolute counts are converted to relative abundances to account for differences in sequencing depth. Specifically, the relative abundance of each bacterial family is calculated by dividing its count by the total count across all bacterial families for each sample ~\citep{weiss_normalization_2017, lin_analysis_2020}. Throughout the real data analysis in this article, all abundance measurements refer to relative abundances, which sum to one across all bacterial families at each time point for each patient, thereby accounting for the compositional nature of the microbiota data.

In the cohort, most bacterial families exhibit zero or near-zero relative abundance in the majority of samples, providing minimal information for network reconstruction. Therefore, we only retain those families whose relative abundances exceeded $0.1\%$ in at least $10\%$ of samples following commonly used preprocessing procedures~\citep{mokhtari_filtering_2022, risely_phylogeny-_2021, nikodemova_effect_2023}. This filtering yields 29 distinct bacterial families in total, of which 21 are common to both the pre-HCT and post-HCT study periods, whereas 4 are unique to the pre-HCT and the other 4 are unique to the post-HCT. Thus, each pre- and post-HCT data comprises $p = 25$ bacterial families, which serve as the nodes of the reconstructed networks. Patients with at least three samples within their respective study periods are included, resulting in $R_{\text{pre}} = 274$ patients for the pre-HCT analysis and $R_{\text{post}} = 719$ patients for the post-HCT analysis. 

For each patient, samples are collected at irregular, sparse, and unequally-spaced time points across patients, reflecting the clinical constraints of sample collection in practice. To recover the initial smooth trajectories from each patient's longitudinal observations, we apply PACE, following Equation~\eqref{eq:pace-r}. Since PACE requires input values to be defined on the real line, a logit transformation is applied to each relative abundance prior to estimation. Additionally, a small positive constant $\delta = 10^{-6}$ is added to shift each observation away from the boundary of $[0,1]$ before applying the logit, ensuring the transformation is well-defined. After the initial smoothing, the inverse logit transformation is applied and $\delta$ is subtracted, returning the smoothed trajectories to the original scale. The analysis is conducted separately before and after the transplant study period.

\subsection{Pre-HCT Regulatory Network}\label{sec:pre}

\begin{figure}[t]
\centering

\begin{subfigure}{4.78in}
\centering
\includegraphics[scale = 0.33]{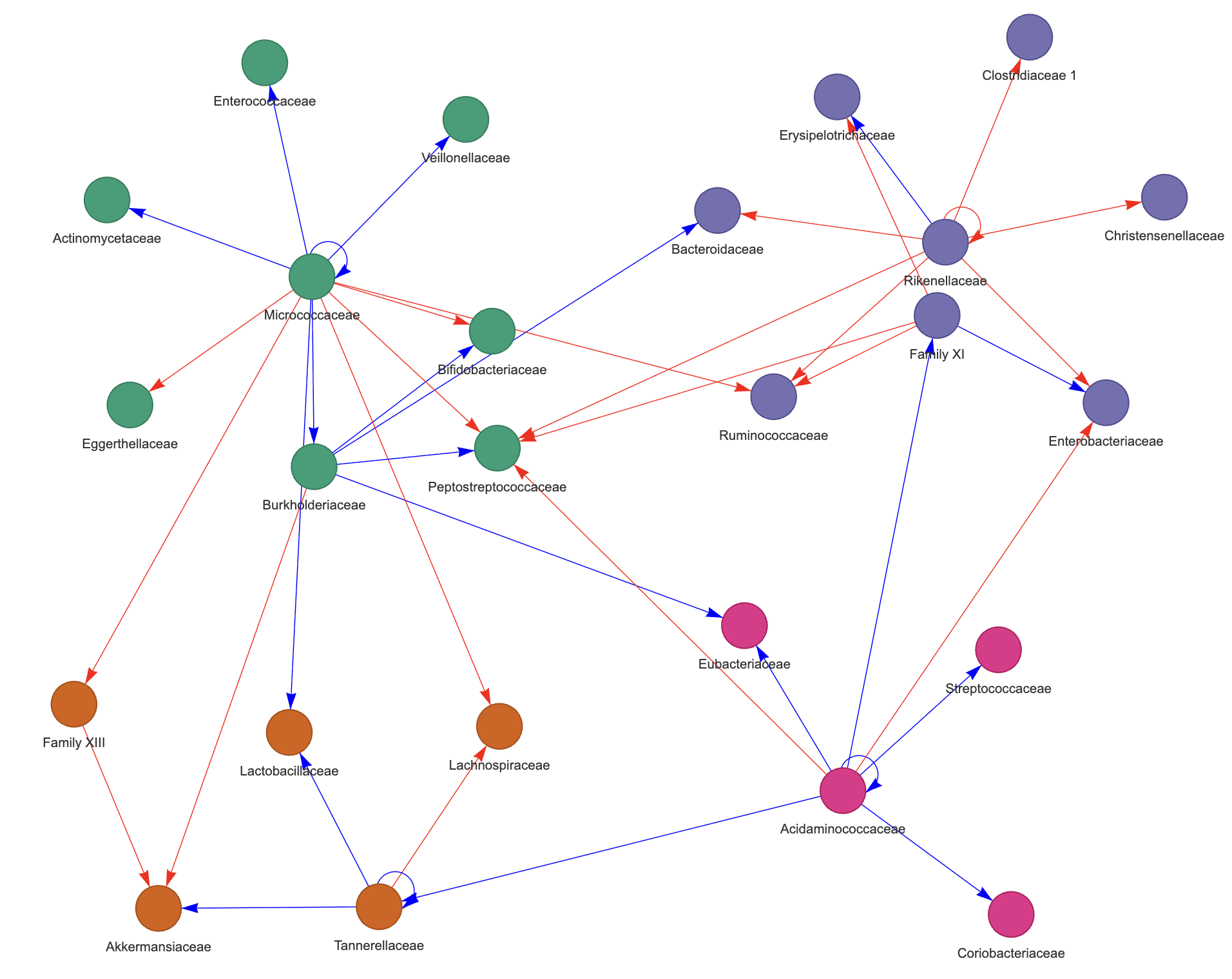}
\end{subfigure}
\hfill
\begin{subfigure}{0.56in}
\centering
\includegraphics[scale = 0.4]{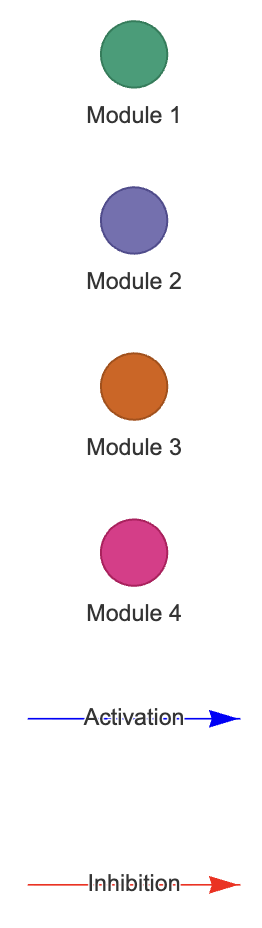}
\end{subfigure}
\caption{Enhanced Causal Omnidirectional Network reconstructed by RECON for the pre-HCT gut microbiota, comprising 25 bacterial families. Each node represents one bacterial family, with node colors indicating module membership identified by the Louvain community detection algorithm. Blue arrows denote activatory regulatory effects, and red arrows denote inhibitory regulatory effects. See Supplementary Figure S2.1 for an interactive version of this plot.}
\label{fig:pre-network}
\end{figure}

\begin{table}[b]
\centering
\fontsize{9}{11}\selectfont
\caption{Network characteristics of the three keystone families in the pre-HCT network reconstructed by RECON.}
\begin{tabular}{llllll}
\hline
Family & Module & Degree & Betweenness & Incoming & Outgoing \\
\hline
\textit{Micrococcaceae} & 1 & 13 & 130.17 &
\begin{tabular}[t]{@{}l@{}}
Activations: 1 \\
Inhibitions: 0
\end{tabular} &
\begin{tabular}[t]{@{}l@{}}
Activations: 6 \\
Inhibitions: 6
\end{tabular} 
\vspace{0.2cm}
\\  [6pt]
\textit{Acidaminococcaceae} & 4 & 9 & 67.02 &
\begin{tabular}[t]{@{}l@{}}
Activations: 1 \\
Inhibitions: 0
\end{tabular} &
\begin{tabular}[t]{@{}l@{}}
Activations: 6 \\
Inhibitions: 2
\end{tabular}  
\vspace{0.2cm}
\\   [6pt]
\textit{Rikenellaceae} & 2 & 9 & 62.14 &
\begin{tabular}[t]{@{}l@{}}
Activations: 0 \\
Inhibitions: 1
\end{tabular} &
\begin{tabular}[t]{@{}l@{}}
Activations: 1 \\
Inhibitions: 7
\end{tabular}
\\
\hline
\end{tabular}
\label{tab:pre-keystones}
\end{table}

The pre-HCT network, displayed in Figure~\ref{fig:pre-network}, comprises $p = 25$ nodes and 42 directed edges, equally divided between activatory and inhibitory interactions (21 each). The network exhibits a modularity score of $H = 0.4113$, indicating a pronounced modular structure~\citep{luo_progress_2024}. Four modules are identified via the Louvain community detection method: Modules 1 and 2 are the largest, each comprising eight families, while Modules 3 and 4 are smaller. Modules 1 and 4 each interact directly with all other modules, whereas Modules 2 and 3 share no direct inter-module interactions with each other. This modular organization is consistent with the ecological principle that communities partitioned into loosely connected modules are more resilient to perturbation, as disruptions within one module are less likely to propagate across the full network~\citep{lima-mendez_determinants_2015, faust_microbial_2012}.

Among the 25 bacterial families, three stand out, exhibiting exceptionally high degrees and betweenness centralities: \textit{Micrococcaceae} (degree $= 13$, betweenness $= 130.17$), \textit{Acidaminococcaceae} (degree $= 9$, betweenness $= 67.02$), and \textit{Rikenellaceae} (degree $= 9$, betweenness $= 62.14$). 
Following the definition of \citet{berry_deciphering_2014}, families with high degrees and high betweenness centralities are considered potential keystone families, as their high connectivity and bridging role suggest that their removal could substantially affect the structure and stability of the entire network. The network characteristics of these three families are summarized in Table~\ref{tab:pre-keystones}. \textit{Micrococcaceae} exhibits the highest degree in the network, with 12 edges comprising one activatory self-edge and 11 outgoing edges (five activatory and six inhibitory) regulating other families distributed across all four modules. Its betweenness centrality of $130.17$ is significantly higher than that of any other family, indicating that it lies on a substantial proportion of shortest paths between node pairs and serves as the primary structural bridge of the pre-HCT community. \textit{Acidaminococcaceae} has one activatory self-edge, and seven outgoing regulatory edges (five activatory, two inhibitory) targeting other families that are in Modules 1, 3, and 4. \textit{Rikenellaceae} has an inhibitory self-edge and seven outgoing regulatory edges (six inhibitory, one activatory) targeting other families. Among all 25 families, \textit{Peptostreptococcaceae} receives the highest number of incoming edges (five), of which four are inhibitory and one is activatory, identifying it as a major target of regulation in the pre-transplant community.

\begin{figure}[htbp]
\centering

\begin{subfigure}{3.3in}

\includegraphics[width=\linewidth]{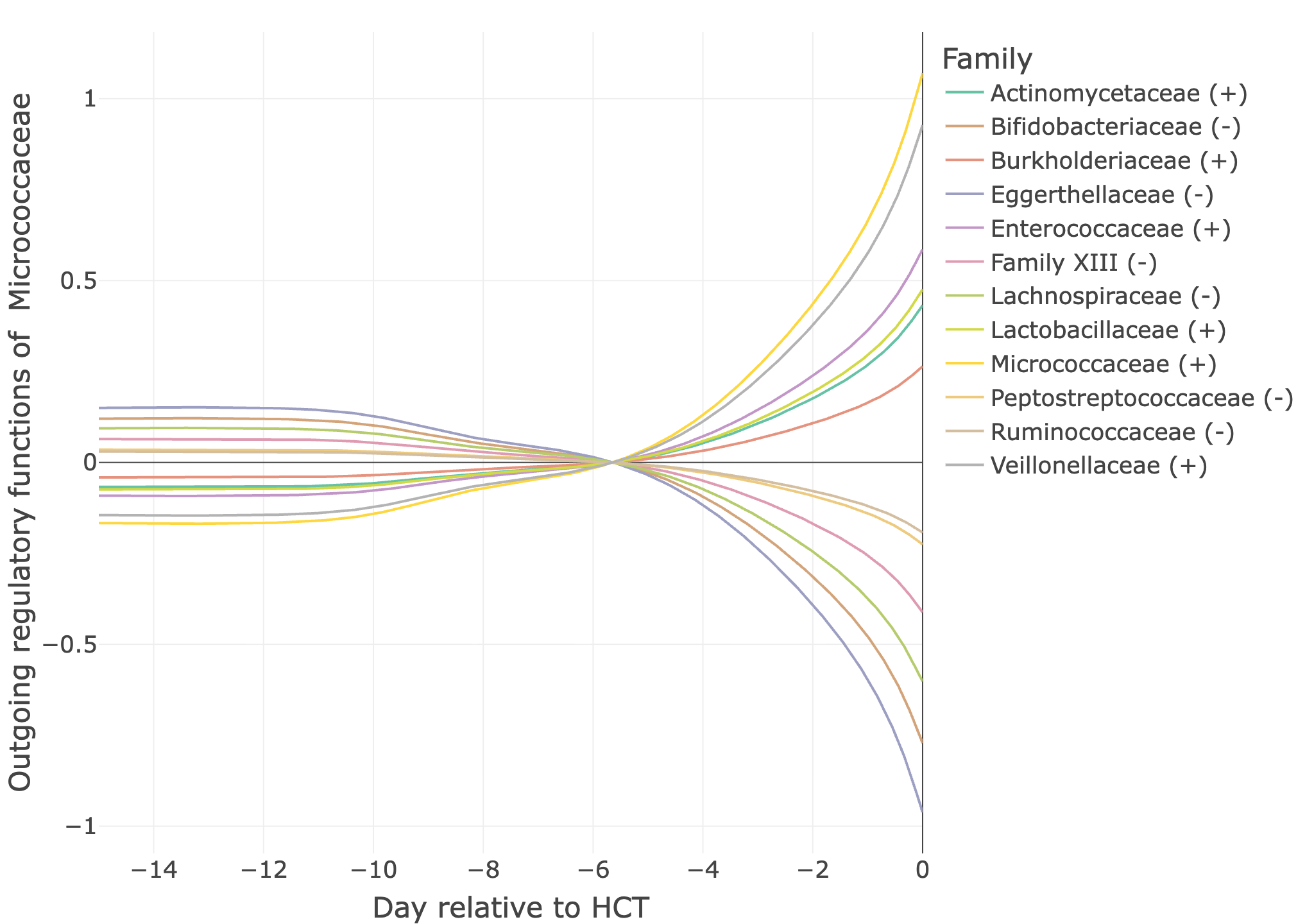}
\caption{Micrococcaceae}
\label{fig:pre-dynamics-micro}
\end{subfigure}
\hfill
\begin{subfigure}{3.3in}

\includegraphics[width=\linewidth]{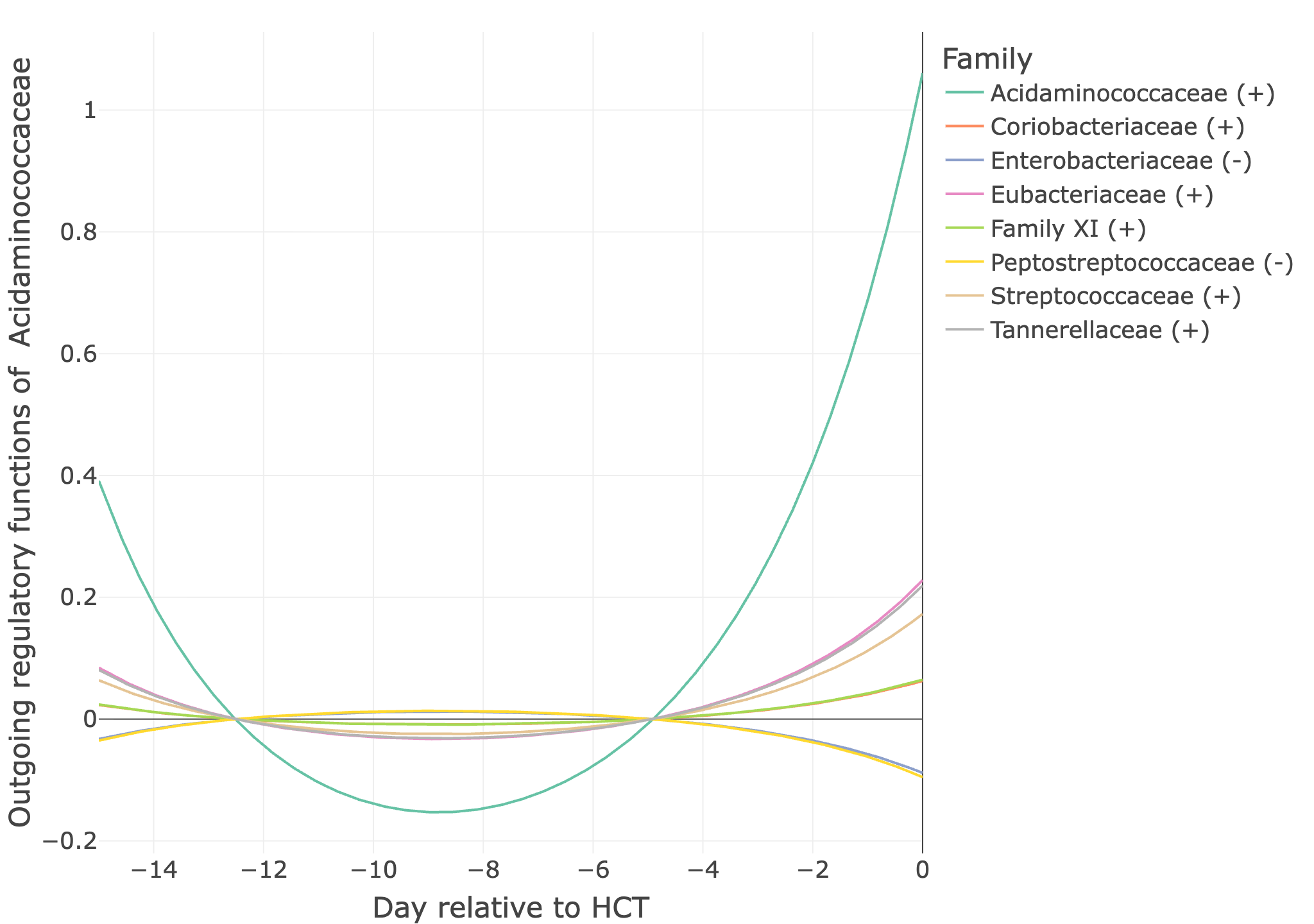}
\caption{Acidaminococcaceae}
\label{fig:pre-dynamics-acida}
\end{subfigure}

\begin{subfigure}{3.3in}

\includegraphics[width=\linewidth]{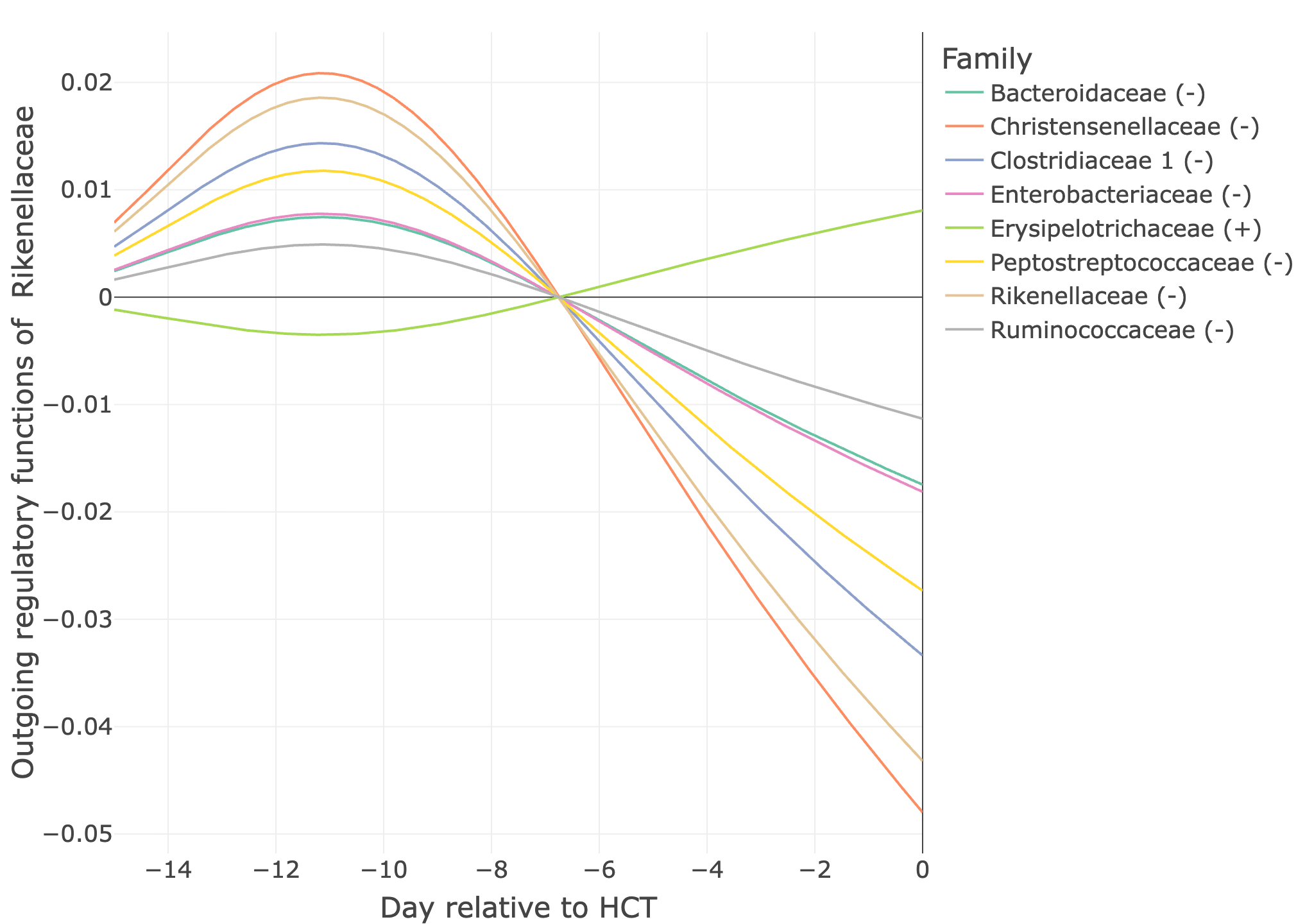}
\caption{Rikenellaceae}
\label{fig:pre-dynamics-rik}
\end{subfigure}

\caption{The estimated outgoing regulatory functions of the three keystone families in the pre-HCT network reconstructed by RECON over the 15 days study period preceding transplantation. The $x$-axis denotes days relative to transplantation, and the $y$-axis denotes the estimated regulatory function, $\widehat{f}_{jk}(\widehat{X}(t;\mathbf{h},\boldsymbol{\theta}))$. Panel (a): keystone family - \textit{Micrococcaceae}. Panel (b): keystone family - \textit{Acidaminococcaceae}. Panel (c): keystone family - \textit{Rikenellaceae}. The legend identifies the target bacterial family being regulated by the keystone families. Positive ($+$) and negative ($-$) values indicate activatory and inhibitory regulatory edge, respectively. See Supplementary Figure S2.2 for an interactive version of this plot.}
\label{fig:pre-dynamics}
\end{figure}

Figure~\ref{fig:pre-dynamics} presents the estimated regulatory functions, $\widehat{f}_{jk}(\widehat{X}_k(t;\mathbf{h},\boldsymbol{\theta}))$, for each outgoing edge of \textit{Micrococcaceae}, \textit{Acidaminococcaceae}, and \textit{Rikenellaceae}, over the 15 days preceding HCT study period, as defined in Equation~\eqref{eq:f-hat}. In each panel, $k$ indexes each of the reference keystone families and $j$ indexes the other families it regulates, so that $\widehat{f}_{jk}$ describes the estimated outgoing regulatory effect of $k \rightarrow j$. These trajectories illustrate how the regulatory influence exerted by each of these three keystone families on other families evolves over time. In Panel~\ref{fig:pre-dynamics-micro}, the regulatory effects of \textit{Micrococcaceae} remain nearly flat from day $-15$ to approximately day $-10$, after which all trajectories converge toward zero, reaching zero near day $-6$. Beyond this point, the effects diverge rapidly as day 0 approaches, producing a fan-out pattern. We can observe that multiple pairs of trajectories appear as close mirror images of each other. Among these mirrored pairs, the activatory effect on \textit{Enterococcaceae} and the inhibitory effect on \textit{Lachnospiraceae} closely mirror each other, as does the inhibitory effect on \textit{Eggerthellaceae} mirror the activatory effect on \textit{Veillonellaceae}, and the activatory effect on \textit{Actinomycetaceae} mirrors with the inhibitory effect on \textit{Family XIII}. In Panel~\ref{fig:pre-dynamics-acida}, the regulatory effects of \textit{Acidaminococcaceae} exhibit a quadratic trend. The self-regulation function displays a substantially larger amplitude than all other functions, forming a concave-up parabola with a trough near day $-9$, before rising steeply toward day 0. The remaining functions are considerably smaller in magnitude, crossing zero twice, once near day $-13$ and again around day $-5$. The activatory effects on \textit{Family XI} and \textit{Coriobacteriaceae} both closely mirror the inhibitory effect on \textit{Enterobacteriaceae}. In Panel~\ref{fig:pre-dynamics-rik}, the regulatory effects of \textit{Rikenellaceae} follow a parabolic trend with peaks and troughs centered near day $-11$, converging toward zero around day $-7$. Following this convergence, the inhibitory curves continue to decrease while the sole activatory effect on \textit{Erysipelotrichaceae} increases, closely mirroring the inhibitory effect on \textit{Ruminococcaceae}.

\subsection{Post-HCT Regulatory Network}\label{sec:post}

\begin{figure}[tb]
\centering

\begin{subfigure}{4.78in}
\centering
\includegraphics[scale = 0.33]{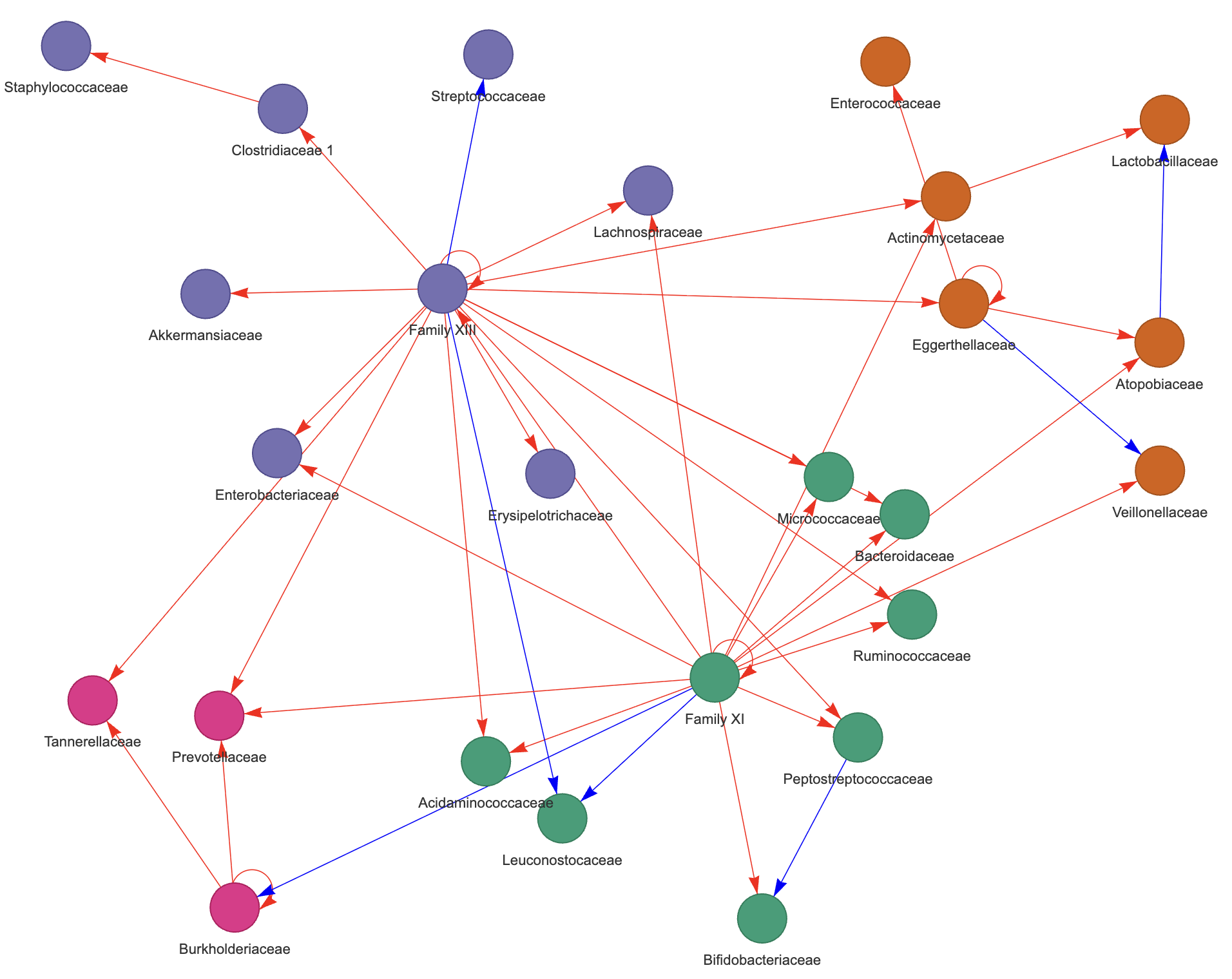}
\label{fig:post-network-image}
\end{subfigure}
\hfill
\begin{subfigure}{0.56in}
\centering
\includegraphics[scale = 0.4]{figures/alloHCT/colors.png}
\end{subfigure}
\caption{Enhanced Causal Omnidirectional Network reconstructed by RECON for the post-HCT gut microbiota, comprising 25 bacterial families. Each node represents one bacterial family, with node colors indicating module membership identified by the Louvain community detection algorithm. Blue arrows denote activatory regulatory effects, and red arrows denote inhibitory regulatory effects. See Supplementary Figure S2.3 for an interactive version of this plot.}
\label{fig:post-network}
\end{figure}

The post-HCT network, displayed in Figure~\ref{fig:post-network}, comprises $p = 25$ nodes and $44$ directed edges, of which $7$ are activatory and $37$ are inhibitory. This composition stands in stark contrast to the pre-HCT network, indicating a marked shift toward a more competitive gut microbiota community following transplantation. The network exhibits a modularity score of $H = 0.2998$, reflecting a substantially weaker modular structure compared to the pre-HCT network ($H = 0.4113$). Four modules are identified via the Louvain community detection method. Modules 1 and 2 are the largest, each containing eight families, while Module 4 is the smallest, containing only three families. Both Modules 1 and 2 interact with all other modules, whereas no direct interactions are observed between Modules 3 and 4.

\begin{table}[t]
\centering
\fontsize{9}{11}\selectfont
\caption{Network characteristics of the two keystone families in the post-HCT network reconstructed by RECON.}
\begin{tabular}{llllll}
\hline
Family & Module & Degree & Betweenness & Incoming & Outgoing \\
\hline
\textit{Family XIII} & 2 & 19 & 167.37 &
\begin{tabular}[t]{@{}l@{}}
Activations: 0 \\
Inhibitions: 2
\end{tabular} &
\begin{tabular}[t]{@{}l@{}}
Activations: 2 \\
Inhibitions: 15
\end{tabular}
\vspace{0.2cm}
\\[6pt]
\textit{Family XI} & 1 & 17 & 89.87 &
\begin{tabular}[t]{@{}l@{}}
Activations: 0 \\
Inhibitions: 1
\end{tabular} &
\begin{tabular}[t]{@{}l@{}}
Activations: 2 \\
Inhibitions: 14
\end{tabular}
\\
\hline
\end{tabular}
\label{tab:post-keystones}
\end{table}

Among the 25 bacterial families, two nodes stand out, exhibiting the highest degrees and betweenness centralities: \textit{Family XIII} (degree $= 19$, betweenness $= 167.37$) and \textit{Family XI} (degree $= 17$, betweenness $= 89.87$). The network characteristics of these two families are summarized in Table~\ref{tab:post-keystones}. In contrast to the pre-HCT network, where betweenness centrality is distributed across several families, $15$ of the remaining $23$ families have zero betweenness centrality in the post-HCT network, reflecting an extreme concentration of bridging influence in the two keystone families. Together, the outgoing edges of \textit{Family XI} and \textit{Family XIII} account for $75\%$ of the total edges in the network. \textit{Family XIII} exhibits the highest degree in the post-HCT network, with $2$ incoming edges and $17$ outgoing edges. It receives no activatory incoming edges and is inhibited by \textit{Family XI} and itself. Its outgoing edges are predominantly inhibitory, inhibiting $15$ other families while activating only \textit{Leuconostocaceae} and \textit{Streptococcaceae}. \textit{Family XI} has one inhibitory self-edge and $15$ outgoing edges, inhibiting $13$ other families while activating only \textit{Burkholderiaceae} and \textit{Leuconostocaceae}. \textit{Family XI} and \textit{Family XIII} jointly inhibit nine families, while \textit{Leuconostocaceae} is the only family activated by both of them. The dominance of inhibitory outgoing edges and the extreme concentration of betweenness centrality suggest that \textit{Family XI} and \textit{Family XIII} serve as potential keystone families in the post-HCT microbiota, exerting competitive regulatory control over the majority of the community. \textit{Prevotellaceae} is the family with the maximum number of incoming edges, all of which are inhibitory.

\begin{figure}[!t]
\centering

\begin{subfigure}{3.37in}
\includegraphics[width=\linewidth]{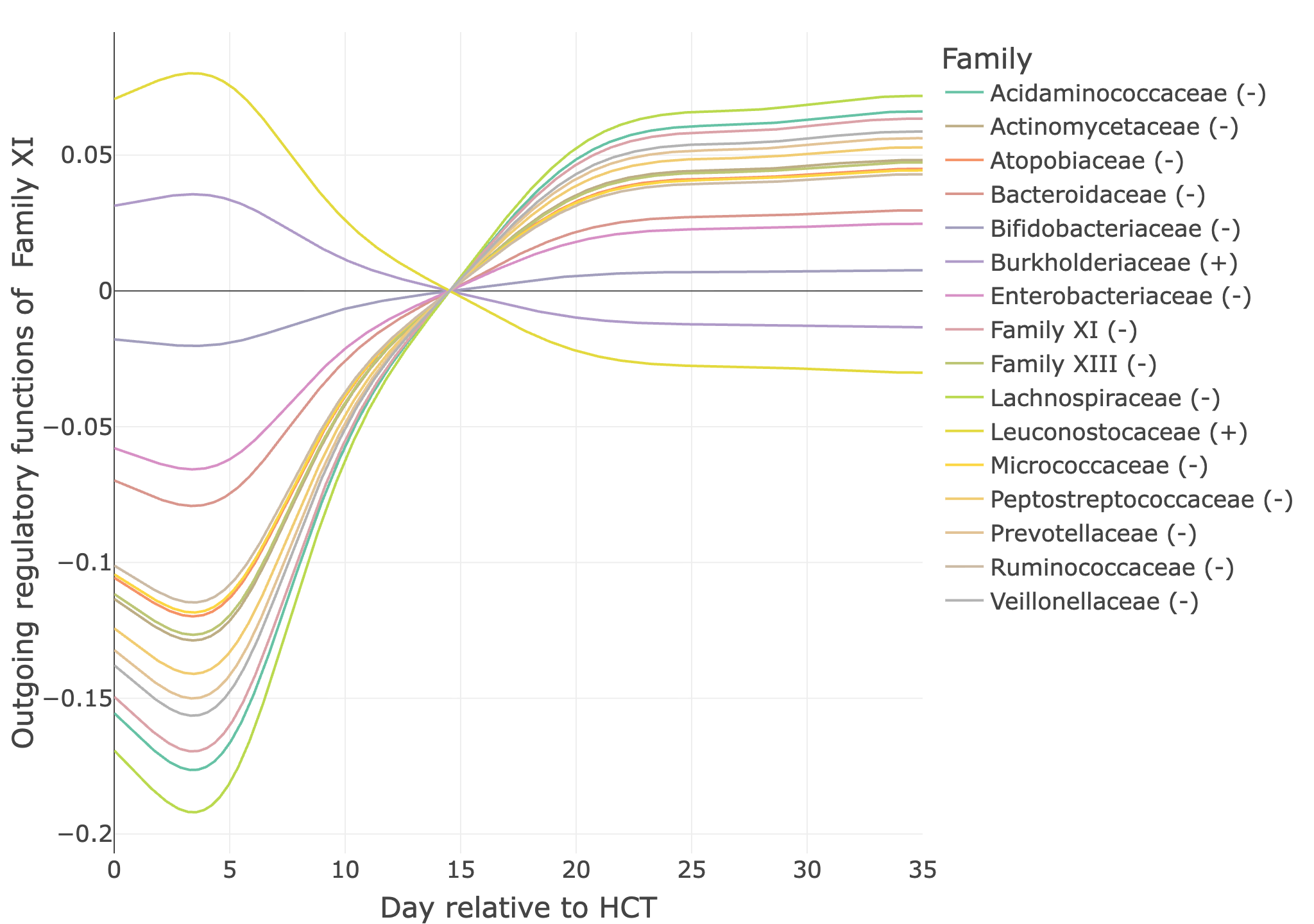}
\caption{Family~XI}
\label{fig:post-dynamics-xi}
\end{subfigure}
\hfill
\begin{subfigure}{3.37in}
\includegraphics[width=\linewidth]{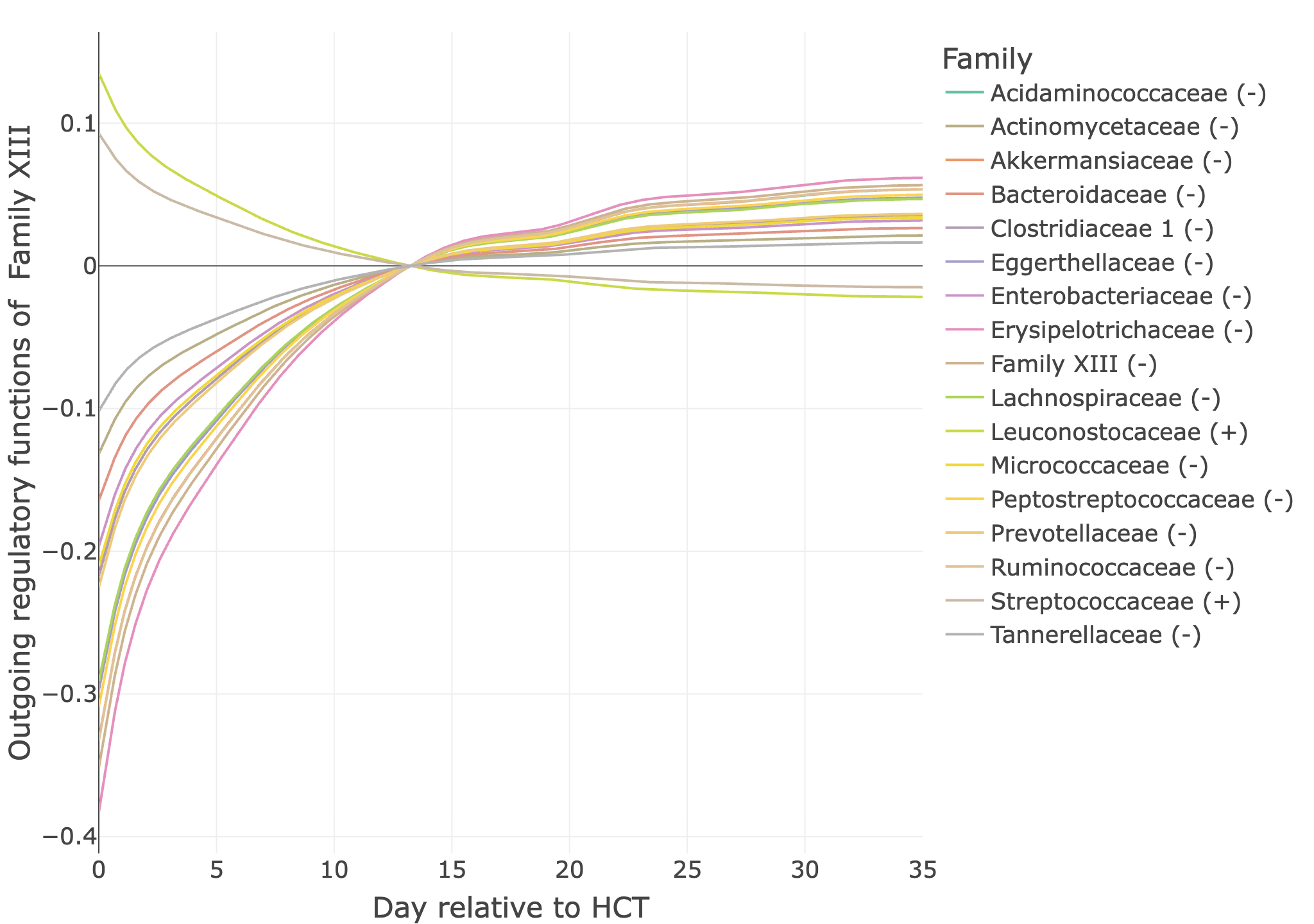}
\caption{Family~XIII}
\label{fig:post-dynamics-xiii}
\end{subfigure}

\caption{The estimated outgoing regulatory functions of the two keystone families in the post-HCT network reconstructed by RECON over the 35 days study period after transplantation. The $x$-axis denotes days relative to transplantation, and the $y$-axis denotes the estimated regulatory function, $\widehat{f}_{jk}(\widehat{X}(t;\mathbf{h},\boldsymbol{\theta}))$. Panel (a): keystone family- \textit{Family XI}. Panel (b): keystone family - \textit{Family XIII}. The legend identifies the target bacterial family being regulated by the keystone families. Positive ($+$) and negative ($-$) values indicate activatory and inhibitory regulatory edge, respectively. See Supplementary Figure S2.4 for an interactive version of this plot.}
\label{fig:post-dynamics}
\end{figure}

Figure~\ref{fig:post-dynamics} presents the estimated outgoing regulatory functions $\widehat{f}_{jk}(\widehat{X}_k(t;\mathbf{h},\boldsymbol{\theta}))$ of \textit{Family XI} and \textit{Family XIII} over the 35 days study period following HCT. In Panel~\ref{fig:post-dynamics-xi}, the regulatory functions of \textit{Family XI} exhibit the greatest curvature between days 0 and 5, after which all functions converge toward zero, reaching zero near day 15. Beyond this point, the regulatory functions change direction and plateau. The regulatory function corresponding to the inhibitory effect on \textit{Lachnospiraceae} is the most pronounced throughout the post-HCT study period, reaching the largest magnitude around day 3 before converging with the remaining curves. The activatory effect of \textit{Family XI} on \textit{Bacteroidaceae} closely mirrors its inhibitory effect on \textit{Leuconostocaceae}. In Panel~\ref{fig:post-dynamics-xiii}, all trajectories exhibit their most pronounced bend near day 2 and converge 
toward zero around day 13. After day 13, the regulatory functions corresponding to inhibitory effects plateau at small positive values ($< 0.1$) and those corresponding to activatory effects plateau at small negative values ($> -0.1$). The inhibitory effect of \textit{Family XIII} on \textit{Actinomycetaceae} closely mirrors its activatory effect on \textit{Leuconostocaceae}, and its inhibitory effect on \textit{Tannerellaceae} closely mirrors its activatory effect on \textit{Streptococcaceae}.

Interactive versions of the regulatory networks in Figures~\ref{fig:pre-network} and~\ref{fig:post-network}, and of the regulatory functions in Figures~\ref{fig:pre-dynamics} and~\ref{fig:post-dynamics}, are provided in S2. The network figures, built with \texttt{visNetwork}, allow users to drag and rearrange nodes, zoom and pan across the graph, click a node to highlight it along with its directly connected neighbors, and search for a specific family by name~\citep{almende_bv_visnetwork_2015}. The function plots, built with \texttt{plotly}, allow users to zoom and pan along both axes, hover over any trajectory to read its exact value at a given time point, and toggle individual trajectories on and off via the legend to isolate and compare any selected subset of the regulatory relationships listed in the legend of the plot~\citep{sievert_plotly_2015}.

\section{Discussion}\label{sec:discussion}

We propose RECON, a new approach to reconstruct regulatory networks from discretely measured time-course state trajectory data under an integral-based nonparametric ODE formulation. Its main methodological contributions lie in five aspects. First, the proposed adaptive edge selection procedure simultaneously leverages group LASSO, Gaussian Mixture Model, and the maximum-ratio criterion to effectively distinguish true regulatory effects from estimation noise, substantially reducing spurious edges while preserving true ones. Second, RECON reconstructs omnidirectional regulatory networks that characterize causal relationships rather than merely statistical associations. Third, by incorporating an initial smoothing and interpolation step, RECON accommodates both dense regular and sparse irregular longitudinal sampling scenarios, thereby considerably broadening the applicability of standard ODE-based network reconstruction. Fourth, RECON models regulatory effects as time-varying functions rather than constant coefficients, yielding a dynamic regulatory network in which both the trajectory of nodes and the regulatory effects evolve over time. This functional representation reveals how regulatory relationships strengthen, weaken, or reverse over time, providing a more realistic description of dynamic biological regulation. Fifth, RECON  enables comprehensive biological interpretation through the two-way direction, sign, and strength of each regulatory edge, distinguishing between activatory and inhibitory regulation, and identifying keystone nodes and the network's topological structure.

The simulation studies demonstrate that RECON substantially improves network reconstruction accuracy across diverse settings and consistently outperforms GRADE, particularly in more challenging scenarios. Across linear and nonlinear ODE systems and varying network structures and sizes, RECON consistently removes nearly all spurious edges while retaining almost all true regulatory edges. This improvement is particularly evident in the most challenging scenario, where the number of spurious edges is reduced from 239 to 0.

In addition, application of RECON to the MSK hospitalome dataset demonstrates its practical utility for uncovering new scientific insights, revealing markedly different network dynamics for the pre- and post-transplant study period. In the pre-HCT network, \textit{Micrococcaceae}, \textit{Acidaminococcaceae}, and \textit{Rikenellaceae} are identified as potential keystone taxa based on their high degree and betweenness centralities. In contrast, \textit{Family XI} and \textit{Family XIII} emerge as potential keystone taxa in the post-HCT network, with predominantly inhibitory outgoing regulatory effects. These taxa correspond to the provisional \textit{Clostridiales} groups \textit{Family XI} and \textit{Family XIII}, historically labeled \textit{Incertae Sedis} because they could not be assigned to formally named families. They are now recognized as the families \textit{Peptoniphilaceae} and \textit{Anaerovoracaceae}, respectively~\citep{johnson_peptoniphilus_2014, whitman_road_2015, chuvochina_proposal_2023}. Their identification as keystone taxa suggests that these recently reclassified bacterial families may play important regulatory roles in post-HCT microbial community dynamics.

Intestinal domination by \textit{Enterococcaceae} occurs frequently in HCT patients worldwide and has been associated with an increased risk of bloodstream infection and HCT treatment failure~\citep{taur_intestinal_2012, holler_metagenomic_2014,stein-thoeringer_lactose_2019,peled_microbiota_2020}. In contrast, members of the \textit{Lachnospiraceae} family, including the genus \textit{Blautia}, have been associated with favorable treatment outcomes in multiple studies~\citep{jenq_intestinal_2015,nguyen_high-resolution_2023}. Our reconstructed post-HCT network suggests that \textit{Family XI} and \textit{Family XIII} indirectly suppress \textit{Enterococcaceae} through \textit{Actinomycetaceae}. To further examine this network-derived hypothesis, we stratify patients according to the abundances of \textit{Family XI} and \textit{Family XIII}. Patient samples with low abundances of both families exhibit a substantially higher relative abundance of \textit{Enterococcaceae}. Together, these findings suggest that \textit{Family XI} and \textit{Family XIII} may help restrain the expansion of \textit{Enterococcaceae} following HCT, thereby potentially contributing to favorable treatment outcomes.

\bibliographystyle{plainnat}
\bibliography{references.bib}

\newpage

\section*{Appendix}

\textit{          }
\setcounter{figure}{0}
\renewcommand{\thefigure}{S\arabic{figure}}

\FloatBarrier

\begin{figure}[H]
\centering
\includegraphics[width=5.62in]{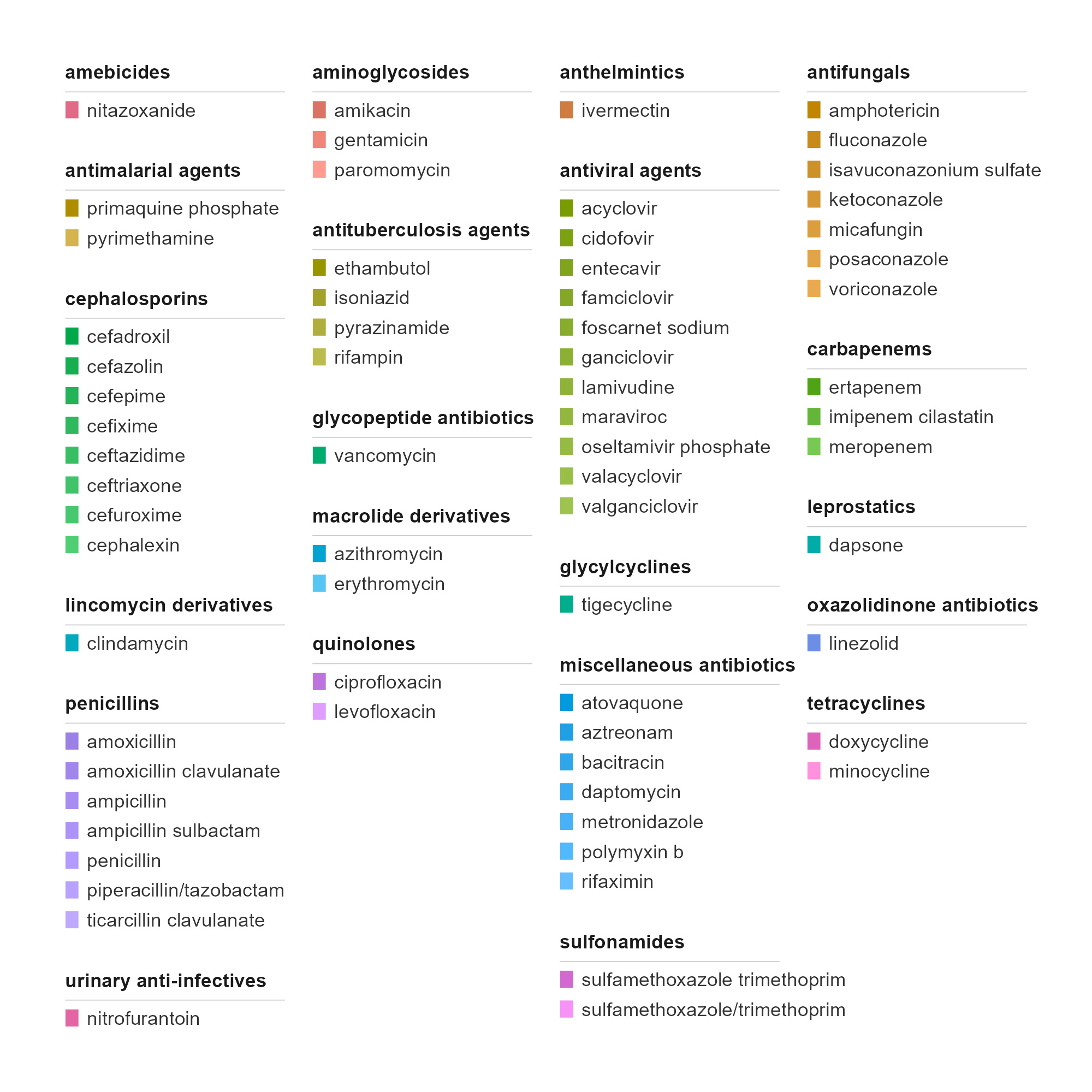}
\caption{The medication information for the legend of Panel (b) of Figure 3. Each medication is assigned a unique color. Medications belonging to the same category are displayed using similar color shades within a common color palette. Medication categories are listed alphabetically.}
\label{fig:drug-legend}
\end{figure}

\setcounter{subsection}{1}  
\renewcommand{\thesubsection}{S\arabic{subsection}}
\subsection{Interactive versions of Figures 9--12}\label{sec:S2}

Click "View" to access each interactive plot from your browser. Each link directs to a page hosted on GitHub Pages \\(\href{https://github.com/knirand1/Supplementary-Material}{github.com/knirand1/Supplementary-Material}). 
\begin{description}
    \item[Figure S2.1: Figure 9] \href{https://knirand1.github.io/Supplementary-Material/Fig_9_Interactive.html}{View}
    \item[Figure S2.2: Figure 10] Panel (a) \href{https://knirand1.github.io/Supplementary-Material/Fig_10a_Interactive.html}{View}, \quad Panel (b) \href{https://knirand1.github.io/Supplementary-Material/Fig_10b_Interactive.html}{View}, \quad Panel (c) \href{https://knirand1.github.io/Supplementary-Material/Fig_10c_interactive.html}{View}
    \item[Figure S2.3: Figure 11]
    \href{https://knirand1.github.io/Supplementary-Material/Fig_11_Interactive.html}{View}
    \item[Figure S2.4: Figure 12] Panel (a) \href{https://knirand1.github.io/Supplementary-Material/Fig_12a_Interactive.html}{View}, Panel (b) \href{https://knirand1.github.io/Supplementary-Material/Fig_12b_Interactive.html}{View}
\end{description}

\end{document}